\documentclass[aip,rsi,amsmath,amssymb,reprint]{revtex4-1}
\usepackage{graphicx}
\usepackage{dcolumn}
\usepackage{bm}
\usepackage{color}
\usepackage[version=3,arrows=pgf-filled]{mhchem}

\begin{document}
\preprint{AIP/123-QED}

\title[Laboratory von H\'amos XAS/XES]{Laboratory von H\'amos X-ray Spectroscopy for Routine Sample Characterization}

\author{Zolt\'an N\'emeth}
\email{nemeth.z@wigner.mta.hu}
\affiliation{Wigner Research Centre for Physics, Hungarian Academy of Sciences, H-1525 Budapest, P.O.B. 49., Hungary}
\author{Jakub Szlachetko}
\email{jszlachetko@ujk.edu.pl}
\affiliation{Institute of Physics, Jan Kochanowski University, 25-406 Kielce, Poland}
\affiliation{Paul Scherrer Institut, 5232 Villigen PSI, Switzerland}
\author{\'Eva G. Bajn\'oczi}
\author{Gy\"orgy Vank\'o}
\email{vanko.gyorgy@wigner.mta.hu}
\affiliation{Wigner Research Centre for Physics, Hungarian Academy of Sciences, H-1525 Budapest, P.O.B. 49., Hungary}

\date{\today } 

\begin{abstract}
High energy resolution, hard X-ray spectroscopies are powerful element selective probes of the electronic and local structure of matter, with diverse applications in chemistry, physics, biology and materials science. The routine application of these techniques is hindered by the complicated and slow access to synchrotron radiation facilities. Here we propose a new, economic, easily operated laboratory high resolution von H\'amos type X-ray spectrometer, which offers rapid transmission experiments for X-ray absorption, and is also capable of recording X-ray emission spectra. The use of a cylindrical analyzer crystal and a position sensitive detector enabled us to build a maintenance free, flexible setup with low operational costs, while delivering synchrotron grade signal to noise measurements in reasonable acquisition times. We demonstrate the proof of principle and give examples for both measurement types. Finally, tracking of a several day long chemical transformation, a case better suited for laboratory than synchrotron investigation, is also presented.
\end{abstract}

\maketitle

\section{Introduction}

Spectroscopies based on the absorption of hard X-rays by core electrons are powerful element specific probes of the local nuclear and electronic structure \cite{Bunker2010,Koningsberger1987,Stern1983}. X-ray absorption near edge structure (XANES) investigates the intensity and position of the features in a 50--100 eV region around the absorption edge to extract information on valence and spin states as well as coordination number. Extended X-ray absorption fine structure spectroscopy (EXAFS) gives a local structural description of the environment of the studied atom, complementary to X-ray or neutron diffraction, with the advantage of element selectivity, but without the requirement of long range order. X-ray emission spectroscopy (XES) complements X-ray absorption spectroscopies (XAS): while XANES traces the lowest-lying unoccupied electronic levels, the fluorescent photons appearing after the core hole excitation by the incoming hard X-rays, which make up the XES spectra, provide information about the occupied electronic states.

Benefiting from these spectroscopic techniques requires a high brilliance hard X-ray source providing sufficient photon flux, combined with a monochromator providing a narrow energy bandwidth necessary to resolve the variations of the absorption coefficient around a specific absorption edge of an element in XAS or the fine structure of the fluorescence spectra in XES. The development of the $3^{\rm rd}$ generation synchrotrons made hard XAS, XES and related techniques available to frontier research\cite{Bunker2010,glatzel2005ccr,deGroot2001}, even at extreme conditions and low concentration.\cite{torchio2014,rueff2010,rovezzi2014} The technique evolved quickly and both XAS and XES have developed into indispensable tools in many disciplines including materials and earth sciences, biology and chemistry.

In addition to frontier research, there is a high demand for X-ray spectroscopies as routine characterization tools in most universities and research institutes. Due to the required high brilliance and energy scanning, X-ray spectroscopy experiments are normally performed at synchrotron beamlines. The exploitation of these techniques is thus hindered by the limited accessibility, and high relative price. The limited number of these oversubscribed beamlines, and the relatively slow proposal review and scheduling system leads to a 6--12 month waiting period. Even industrial customers, with sufficient budget to purchase the expensive beamtime, have a long wait before their experiment can be scheduled. Many modern research projects cannot wait months for simple answers, and thus X-ray techniques are often ignored. Moreover, synchrotrons provide only short access periods (typically a few days), which makes long-term studies hardly possible. Thus, chemical reactions or biological processes with time scales longer than a couple of days are also excluded. Furthermore, the high X-ray flux provided by modern synchrotron sources often exceed the requirements for routine samples, and therefore an otherwise important experiment wastes precious resources. Finally, another disadvantage of the dependence of XAS and XES methods on  synchrotrons is that their rare accessibility does not favor the involvement and proper training of the new generation of X-ray scientists/students, and it is even more difficult to popularize it among a wider user community.

Traditional laboratory XAS has been very time consuming due to the low brilliance of radiation from X-ray tubes. However, advances in monochromatization, detection technology, and sources offer new convenient and affordable solutions for assembling such an instrument. In a recent pioneering work G.\ Seidler and coworkers \cite{Seidler2014,MortensenarXiv2015,SeidlerarXiv2015} have reported a new spectrometer that allows laboratory XAS and XES investigations. It is based on a 1m-diameter Rowland-circle spectrometer, where a spherically-bent Johann analyzer crystal serves as a monochromator, which selects a single wavelength from the radiation of a bright microfocus X-ray source and focuses it on a sample placed in front of a Silicon drift diode (SDD) detector. During acquisition, the X-ray source, the crystal and the detector are all translated to maintain the Rowland circle geometry for each wavelength during the energy scan. The inner space is filled with He to reduce air absorption in the almost 2m path inside the spectrometer. The results confirmed that XAS and XES spectra can be indeed acquired with sufficient count rates in acceptable time frames with the same resolution as is the case with synchrotrons \cite{Seidler2014}. The presented the K absorption edge of a Co foil were taken for 80s/point, which would require ca.\ 6.5 hours net acquisition time for a 300 point XANES spectrum. A complete K$\beta$ XES spectrum of a CoO sample requires about the same acquisition time. While the performance of this apparatus is respectable, it's operation requires simultaneous precise movements of the source and sample, analyzer, and detector. This might render the experimental setup less flexible, not compatible with complex sample environments, and also requires maintenance. Moreover, the instrument is also relatively bulky, and requires a substantial amount of expensive He gas for the operation.

Recently, table-top high-resolution X-ray spectrometers using a different working principle have also been realized, which are based on a laser-powered plasma X-ray source and an energy-dispersive cryogenic microcalorimeter detector.\cite{uhlig2013,joe2016} Such instruments offer the possibility of ultrafast time resolved XAS and XES studies in the laboratory. However, their energy resolution lags behind those of the wavelength dispersive spectrometers. The limitations on the overall detected count rates can also be disadvantageous. Moreover, this type of setup will hardly be available to a wide community due to the large construction and maintenance costs.

We propose an alternative approach using a spectrometer based on cylindrical analyzer crystals in the von H\'amos geometry \cite{szlachetko2012rsi}. These crystals are bent perpendicular to the desired X-ray propagation plane to focus as many photons as possible onto the detector, but flat in the parallel direction to map the energy dispersed spectrum to different detector positions. The von H\'amos geometry, which is used more and more at synchrotrons and X-ray free electron lasers,\cite{szlachetko2012rsi,alonsomori2012,hoszowska2004}, has rarely been used in laboratory based hard X-ray spectrometers. The first examples for dispersive-mode laboratory X-ray spectrometers adapted flat single crystals to monochromatize the white X-ray radiation of common sources\cite{Lecante1994,Inada1994}, but the lack of the focusing bending of the von H\'amos type analyzers in these cases hamper both count rate and resolution. C.\ Schlesiger \textit{et al.}\ built an XAS spectrometer based on a HOPG mosaic crystal.\cite{Schlesiger2015} Although it could produce both XANES and EXAFS spectra of a Ni foil, it is optimized for lowering the necessary acquisition time but cannot provide enough resolution to resolve the fine structure of the XANES spectra. Y.\ Kayser \textit{et al.}\  assembled recently a von H\'amos spectrometer for XES with a 0D SDD detector, which can track or scan a specific energy from the analyzer, but does not exploit the potential of the von H\'amos geometry.\cite{Kayser2014} 

The herein proposed instrument uses a single Si crystal (which can be upgraded to e.g. Ge) analyzer to obtain high energy resolution. In order to make it more economical, instead of bending, a segmented analyzer was used. This has been prepared by cutting narrow strips from a single crystal wafer, and the resulting set of crystals are glued to a cylindrical substrate of 250\,mm radius \cite{szlachetko2012rsi}. This projects a wavelength-dispersed spectrum to the detector, which should be a spatially resolving, 1- or 2-dimensional pixel detector. As a spectrum range spanning several hundred eVs is collected simultaneously, there is no need for scanning, thus the whole setup is fixed during the acquisition for a given energy region, for example the XANES region around a given elemental absorption edge. Furthermore, the spectrometer is relatively small, does not require He or vacuum, and can operate even with conventional X-ray tubes at relatively small power. Easily exchangeable fixed setups can be prepared for different edges, with crystal and detector mounts arranged on portable breadboards. Alternatively, such mounts can be motorized to drive the crystal and detector into appropriate positions for different edges. 

This spectrometer can easily give fast feedback on valence- and spin state as well as local geometry selectively for a wide range of elements with a very moderate operational cost. The spectrometer is easy to operate and doesn't rely on scanning motors or special environments. The von H\'amos method has the advantage of measuring the full spectra at once, while the Johann spectrometer, which works with scanning spherically bent crystals, focuses only a single wavelength, a clear advantage at low count rate experiments. In this sense the two setups are complementary, and not competitive. Typically, the count rate in a given channel between the two setups (using 25\,cm radius segmented crystals for the von H\'amos vs. a 1\,m radius Johann crystal) differ by a factor of 20. When many channels ($>$20) need to be scanned, and the signal to noise ratio is satisfactory, von H\'amos is more advantageous. Many experiments with low count rates and bad S/N conditions would not have been possible with a von H\'amos setup in a reasonable time (e.g. refs \onlinecite{vanko2010,March2015}). On the other hand, in most cases with high count rate and sufficient S/N, which seem to include laboratory XANES, the experiments can be substantially faster with the von H\'amos than with Johann.

In this paper we outline the new setup based on the above-mentioned approach: a quick response and cost-effective high resolution von H\'amos type laboratory X-ray spectrometer capable of recording either XAS or XES spectra. After describing the necessary components and the actual setup, we test the working method with an energy sensitive point (0D) detector and analyze potential distortions of the spectral shape. Afterwards we show examples for a K-edge XANES spectra of Ni and Co based compounds, demonstrating the chemical sensitivity of the method. We also describe an example for tracking a long chemical reaction with the case of the hydrolysis of a Ni(IV) compound. Finally, we demonstrate the capability of XES spectroscopy with K$\alpha$ and K$\beta$ spectra of CoO and the K$\alpha$ spectrum of a Ni-foil.

\section{Experimental}

The three main components of the laboratory X-ray spectrometer are a fixed X-ray tube, a von H\'amos type cylindrical analyzer crystal and a position sensitive detector.

A conventional water-cooled Seifert DX-Cu 12x0.4-s X-ray tube was used for this experiment. The electron beam is focused onto a copper anode, with dimensions of $12 \times 0.4 \, \rm{mm}^2$. In the so-called point focus geometry, which was exploited in the present work, the emitted X-rays are taken by a window at $6^{\circ}$ elevation angle, with the projected source size of $1.2 \times 0.4 \, \rm{mm}^2$ (focusing and dispersing directions, respectively). The radiation at 25\,cm from the tube was measured to have a quasi-circular shape with a diameter of ca. 33\,mm. The X-ray tube was operated at 10\,kV voltage and 40\,mA (XAS) as well as 60\,mA (XES) currents for all measurements. The X-ray flux at the exit window of the tube was measured using a Silicon Drift Diode (Amptek) detector. The X-ray tube was set to 10\,kV and 2\,mA (the lowest working current), and the flux was further attenuated by a 1.56\,$\mu$m thin Al foil. Applying the corresponding corrections, the estimated flux extrapolated for 40\,mA tube current is 1.5 $\times$ 10$^9$\,photons/(s$\times$mm$^2$) for a 300\,eV wide Ni K edge absorption spectrum, e.g. shown in Fig.\ \ref{fig_Nichem}A, or 5.2 $\times$ 10$^9$\,photons/(s$\times$mm$^2$) for photons with energy between the absorption edge of Ni (8333\,eV) and the 10\,keV upper limit of the present XES measurements, using 60\,mA tube current.

The von H\'amos analyzer used was composed of $50 \times 1 \, \rm{mm}^2$ segments of 300\,$\mu$m thin Si(111) wafers, arranged on a cylindrical substrate with 250\,mm bending radius, with total crystal size of 50\,mm $\times$ 100\,mm (dispersion $\times$ focusing).

A position sensitive 1 dimensional Dectris Mythen 1K detector was used to record the dispersed X-ray photons during the experiments. The detector consists of 1280\,pixels with 50\,$\mu$m pixel size in the dispersive direction and 8\,mm height. The threshold was set -- if not indicated differently -- to 8\,keV and 5\,keV for the Ni and Co K edges, respectively, and a 2\,mm high lead-foil slit was placed in front of the detector to match the focus size of the segmented analyzer, and thus reduce background. Energy was calibrated using EXAFS standard metal foils and additionally appearing K$\alpha$ fluorescent peaks (Cu radiation in the case of the Co XAS spectra, Fe K$\alpha$ lines from Si(333) reflections in the case of Ni XAS spectra). Alternatively, an energy sensitive Amptek X-123 0D Silicon drift detector (SDD) was also used to measure the energy spectrum of the radiation at the detector position. The active detector surface is 25\,$\rm{mm}^2$. The energy of the 0D SDD detector was calibrated with a radioactive \ce{^{57}Co} source in a Rh matrix, using its emitted Fe K$\alpha$ and Fe K$\beta$ radiation, as well as the 14.412\,keV nuclear transition of the nucleogenic \ce{^{57}Fe} and the Rh K$\alpha$ fluorescence.

The samples used for the transmission mode XAS experiments were measured in the form of a 13\,mm diameter pellet stuck between two pieces of 25\,$\mu$m Kapton tape. All these samples were mixed with cellulose as a solid solvent except the \ce{K2NiF6}, which was mixed with boron nitride to avoid too fast decomposition. The composition of the pellets was optimized to get an ideal edge jump of 1 in $\Delta\mu_x$ and was calculated by the ABSORBIX (v3.2) module of the MAX software package\cite{Alain2009}. During the measurements the samples were placed right in front of the exit window of the X-ray tube. All the measurements were carried out in air atmosphere. For the X-ray emission measurements, pristine samples of the compounds were pressed into pellets and positioned as described in Section \ref{sec:xes}.

\section{Spectrometer setup}

\subsection{X-ray absorption spectrometer}

\begin{figure}
 \includegraphics[width=\columnwidth]{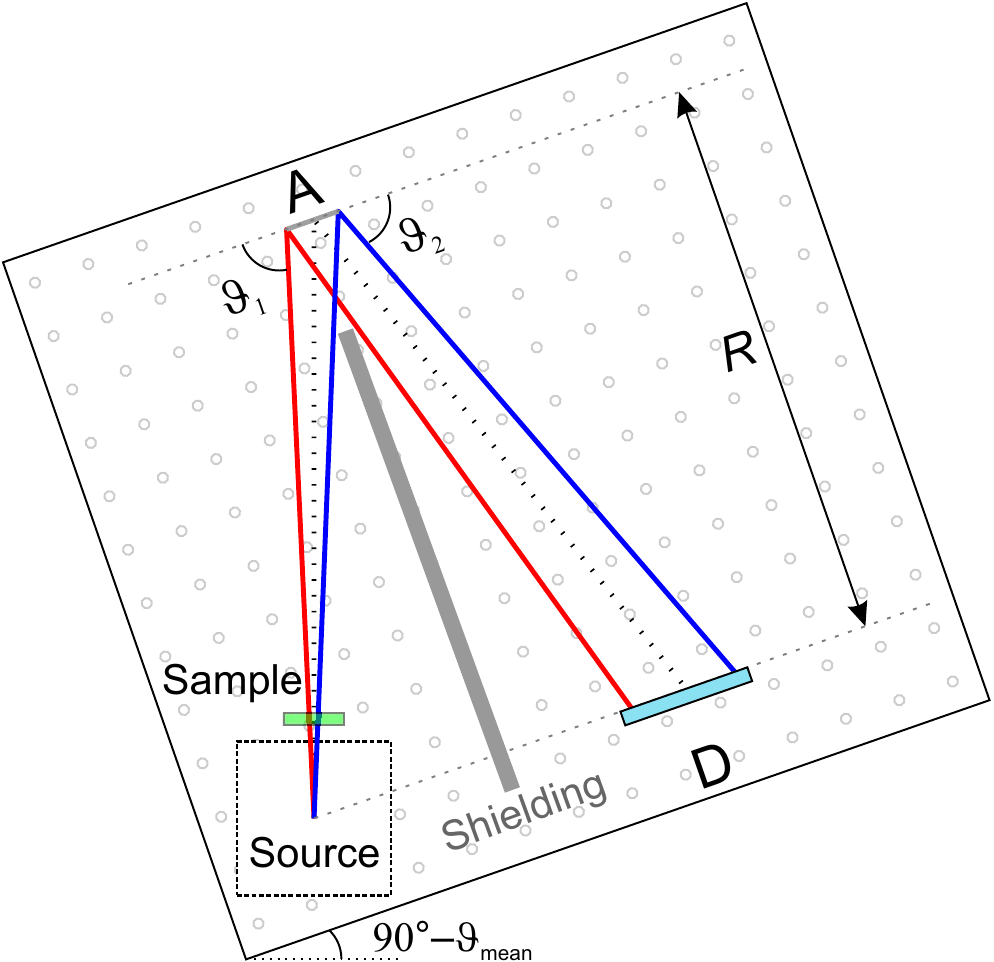}
\caption{(Color online) Top view of the setup of the von H\'amos XAS spectrometer. Two parallel linear translation stages, fixed $R=250$ mm apart on a breadboard carries the (adjustable) mounts for the segmented cylindrical analyzer crystal (A) and the utilized strip detector (D). The breadboard can be rotated around a screw positioned exactly below the X-ray source, so that the analyzer is in the middle of the X-ray beam. The main part of the lead shielding is indicated between the source and the detector as a gray rectangle.}  \label{fig_spectro}
\end{figure}

\begin{figure}
 \includegraphics[width=\columnwidth]{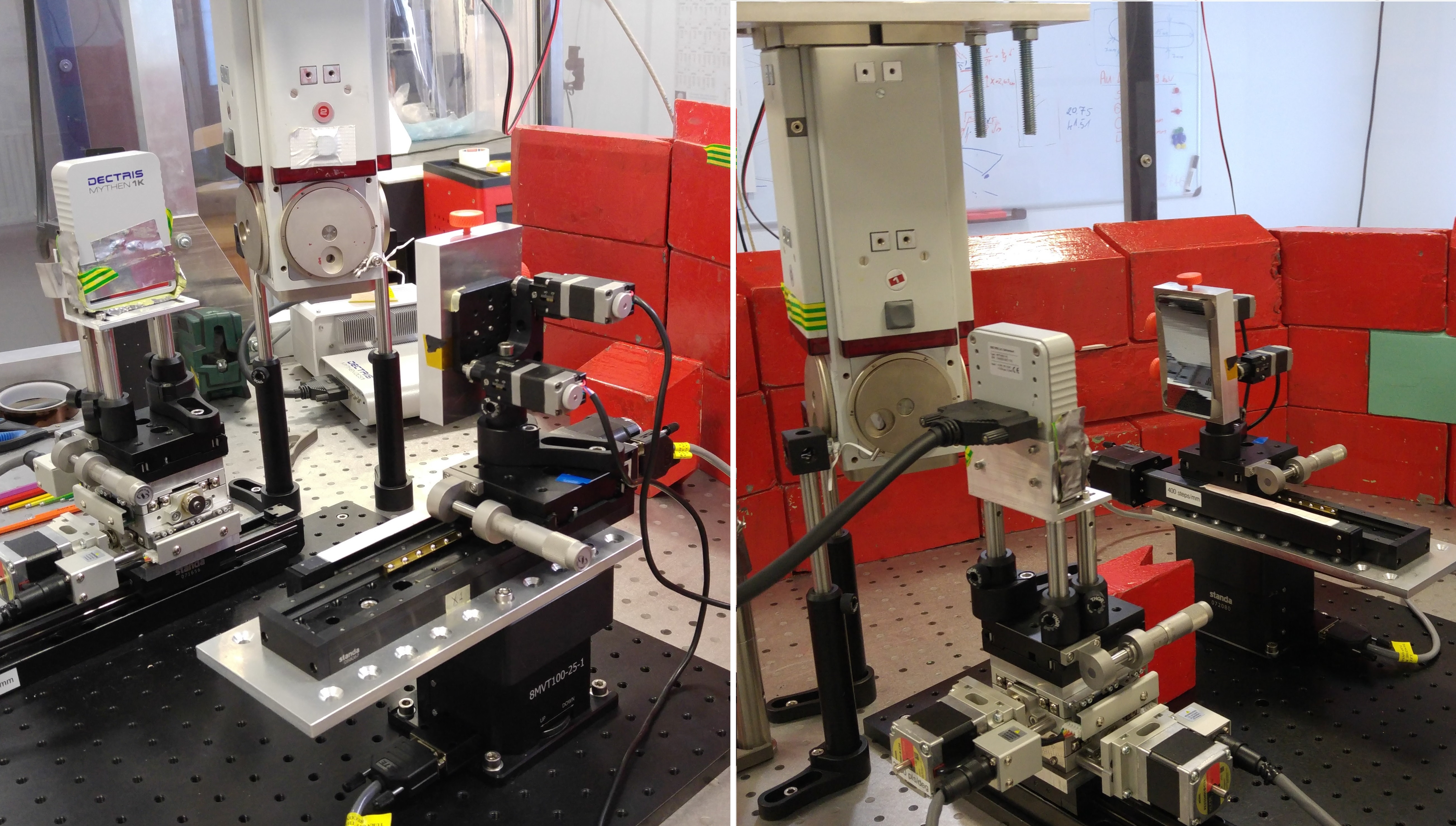}
\caption{(Color online) An actual absorption measurement setup of the spectrometer without the shielding.}  \label{fig_photo}
\end{figure}

The configuration of the X-ray absorption spectrometer is depicted in Fig.\ \ref{fig_spectro}. Both the analyzer crystal and the detector are mounted on a common breadboard, which is partially fixed in a way that in order to be able to optimally position the analyzer crystal in the X-ray beam, it can be rotated with a rotation center exactly below the fixed X-ray source. The positions of the analyzer and detector are calculated to fulfill the Bragg-diffraction of the investigated energy range and the $R$ = 250\,mm focusing radius of the crystal (cf. red and blue lines in Fig.\ \ref{fig_spectro} simulating the propagation of photons with lower and higher energy, respectively). The crystal and the detector are on a 3-axis positioning stage with the possibility for tilting or rotation. Most importantly, the setup includes two long parallel linear translation stages separated by $R$, which, after the initial alignment, are the only ones needed to change the measured energy range (to extend it, or to reach a different edge). This setup allows a several hundred electronvolt wide energy range to be recorded without the need to move any part of the spectrometer. The sample can be placed anywhere in the beam path, however, the optimal position is the closest to the source. The spatial inhomogeneity in the effective thickness of the sample can introduce unwanted intensity variations in the transmitted spectrum, thus the smaller the area of the sample irradiated, the better. Moreover, as the present setup measures only an energy range of a few hundreds of electonvolts, the absorption edge jump cannot be normalized exactly to unity. In order to achieve the best signal to noise ratio, appropriate shielding is necessary to prevent the measured spectrum from being contaminated with unwanted scattered X-rays. The actual setup for absorption measurement is shown in Fig.\ \ref{fig_photo} without the lead shielding.

\subsection{X-ray emission spectrometer}
\label{sec:xes}

In the case of the X-ray emission spectroscopy setup, the source is no longer the anode surface in the X-ray tube, but the excited part of the sample. The Bragg geometry will then be calculated from this origin for the positions of the analyzer and the detector. For the effective recording of the fluorescent photons, the analyzer is best situated perpendicular to the irradiating X-ray beam and the sample is rotated 45$^\circ$ to face both the X-ray tube and the crystal. As the diverging radiation of the X-ray tube excites the sample on a several mm wide surface, the source for the XES becomes big, decreasing the resolution to an unacceptable level. Thus, the sample emission has to be slitted down to have a sub-millimeter projection into the direction of the analyzer.

\section{Results}

\begin{figure}
\includegraphics[width=\columnwidth]{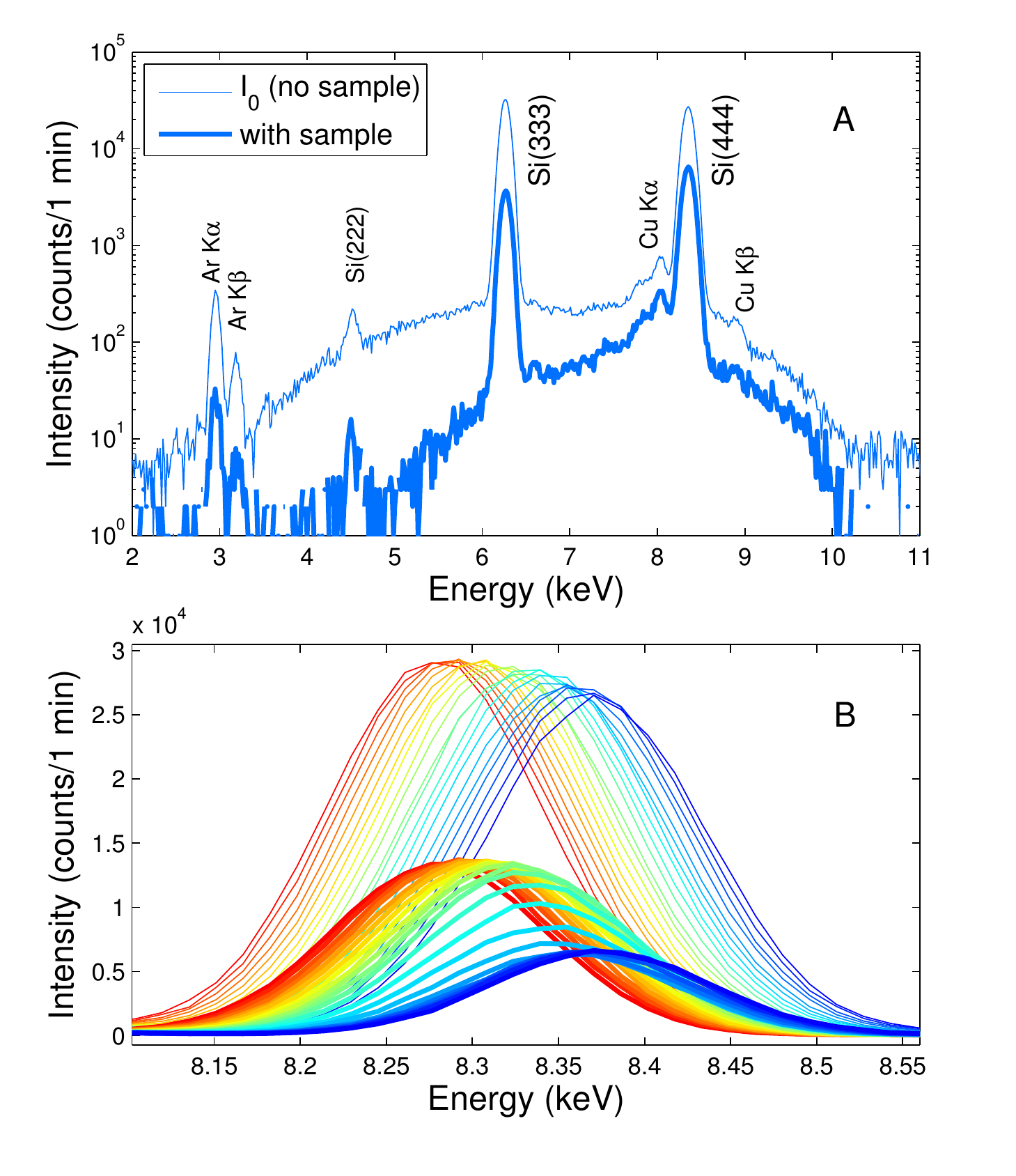}
%
\caption{(Color online) 
\textbf{A:} Energy spectra of the X-ray radiation from the Seifert X-ray tube with and without a NiO sample (thick and thin curves, respectively) reflected by the Si(111) analyzer crystal and recorded by the Amptek 0D detector placed in the detector position described in Fig.\ \ref{fig_spectro}.\\  
\textbf{B:} Dispersion of the Si(444) peak during a detector position scan along the dispersive plane, from the low- to high energy side (direction from red to blue in Fig.\ \ref{fig_spectro}. Thin and thick lines denote again spectra recorded with and without NiO).} \label{fig_Amptek}
\end{figure}

\subsection{Correlation of detector position and energy}  

First we explored the energy spectra of the X-ray radiation diffracted by the crystal and recorded at the detector position along the dispersive plane with the energy-dispersive SDD placed in the focus of the analyzer. The measurements were performed with and without a NiO absorber in the sample position and with the analyzer crystal aligned to diffract energies around the Ni K-edge. The energy spectrum, shown in Fig.\ \ref{fig_Amptek}A, reveals a very wide reflected Bremsstrahlung background, from which the Si(111) crystal reflects intense narrow bands through its diffraction harmonics. Additionally, fluorescence from the Cu tube and the Ar in the air contributes to the background. Translating the 0D detector along the dispersive plane, the Si(111) harmonic peaks drift in energy, as expected from the working principle of the von H\'amos spectrometer. This is shown in Fig.\ \ref{fig_Amptek}B for the Si(444) reflection and X-ray energies around the Ni K-edge. Without an absorber, the reflected intensity is almost flat, this corresponds to the $I_0$ of the XAS.

\begin{figure}
\includegraphics[width=\columnwidth]{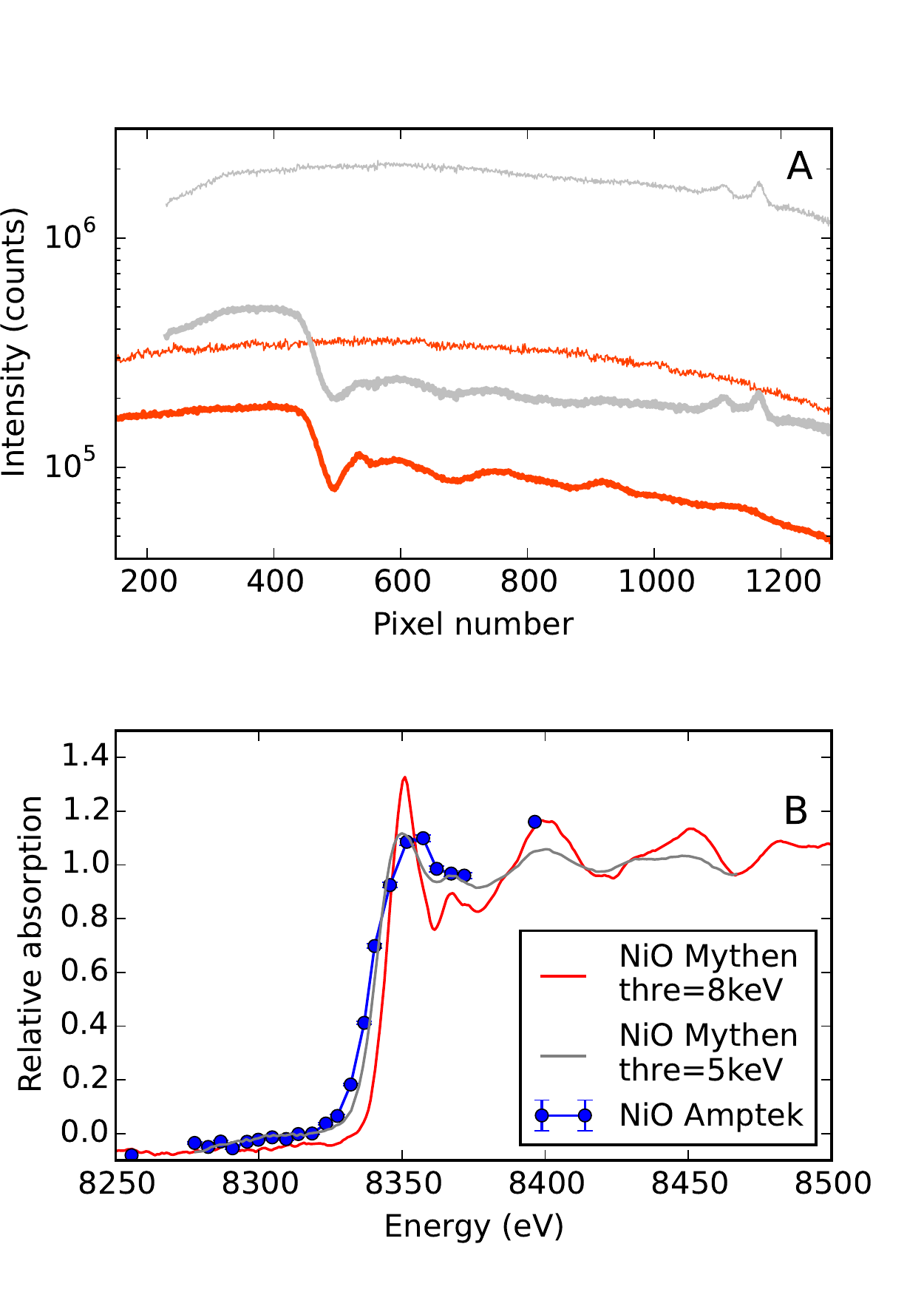}
\caption{(Color online) \textbf{A:} Projected transmission signal on the Mythen detector with and without a NiO sample using 8\,keV detector threshold (red thick and thin curves, 120 and 130\,min acquisition times, respectively). The gray curves present the same scans with 5\,keV threshold (see text). \textbf{B:} Comparison of the XANES of a NiO pellet measured with a fixed Mythen position sensitive detector (red and gray curves) and with scanning an Amptek point detector along the focal line of the dispersed spectrum (blue dots). For the former, the spectrum is binned for every 5 pixel.} \label{fig_NiO}
\end{figure}

With a NiO sample in the beam, a drop in the detected intensity is noticed with additional intensity variations when the diffracted energy is higher than the Ni K-edge, as seen on the lower intensity peaks in Fig.\ \ref{fig_Amptek}B. Indeed, from the logarithm of the ratio of measured intensities the energy dependence of the absorption can be plotted, and it unveils the absorption edge, as is shown in Fig.\ \ref{fig_NiO}B as dots. Although the procedure described above is useful to characterize the spectrometer, it is not a practical method to measure the absorption spectra. Moreover, the big surface of the Amptek sensor decreases the energy resolution considerably, as seen in Fig.\,\ref{fig_NiO}B. The Mythen detector, on the other hand, can measure the intensity simultaneously at 1280 detector positions with 50\,$\mu$m pitch size. Fig.\,\ref{fig_NiO}A shows the Mythen-detected spectra again with and without the NiO sample, as reflected by the von H\'amos Si(111) analyzer crystal (at 69.7$^\circ$ mean Bragg-angle) with two different detector threshold settings, while Fig.\ \ref{fig_NiO}B compares the normalized XANES spectra of the same sample measured with both detectors. In Fig.\ \ref{fig_NiO}A two sets of $I_0$ and NiO transmission spectra is shown: one with detecting in a wide energy window using a low (5\,keV) threshold and one with a narrow energy window (8\,keV threshold), indicated as gray and red curves, respectively. The former settings let the detector capture about one order of magnitude more photons, while the latter may cut in the number of photons even in the energy range of interest. However, the fine structure is clearly damped with the low threshold due to the high background, a relevant part of which originates from the 3rd harmonics contribution (cf. Fig.\ \ref{fig_Amptek}A). Unless one uses an SDD grade discrimination, the role of this background contribution remains an important issue, as analyzed in section \ref{background}. (In the low-threshold data a pair of peaks is visible between pixels 1100 and 1200, which stem from the iron K$\alpha_{1,2}$ fluorescence lines via the Si(333) reflection and arising from X-ray tube anode impurities. This is gone when the threshold exceeds 6.4 keV.) 

The energy range covered by the setup is determined by the size of the Mythen detector along the dispersive axis (provided that the X-ray beam spot is sufficiently large on the analyzer). At the Ni K-edge with Si(444) diffraction and for the length of the detector of 64\,mm, an energy range of about 320\,eV is obtained. Since the detector consists of 1280 pixels with a 50\,$\mu$m pixel size, the sampling intervals on the energy axis are calculated to be about 0.25\,eV/pixel. In order to utilize the same setup for EXAFS measurements, a longer energy range will require data collection at different $\theta_{\rm mean}$ angles, which requires translating the analyzer and the detector to a set of different positions, and turning the base plate accordingly. In this case, sample inhomogeneity, differences in collection geometry and detector (pixel) sensitivity have to be carefully investigated, as similar to data acquisition at synchrotrons, any systematic errors can compromise the data quality.

The spectrometer resolution is determined as the convolution of several factors. First the X-ray diffraction crystal introduces broadening at the level of the Darwin width, which is about 0.1\,eV. Since we employed a segmented-type crystal for the X-ray diffraction, we do not expect additional broadening, as is observed for bent-crystals where the strain causes significant degradation of the resolution. Second, the spatial resolution of the detector will contribute to the total energy resolution by about 0.25\,eV. The third factor is the source size in the dispersive plane. As used in the experiments, the X-ray tube is characterized by a 400\,$\mu$m spot size, which at the Ni K-edge energy and Si(444) diffraction gives rise to about 2\,eV broadening. The total resolution is calculated as the convolution of these three factors, and therefore the setup resolution is estimated to be 2\,eV, primarily due to the effect of the employed source size. Since the spectral sampling of 0.25\,eV is much smaller than the energy resolution of the setup, the measured data can be binned by a factor 4-6 without introducing any further loss of energy resolution to spectral features, resulting in a spectrum with 1-1.5\,eV step size. Should the experiment require, the energy resolution can thus be increased by using a microfocus X-ray tube with a source size of some tens of micrometers or slitting the available X-ray radiation.

With the present setup, in about 2 hours more than $10^5$ counts per pixel can be achieved for $I_0$ even with the higher (8\,keV) threshold, which clearly shows that the experiment is not count-rate limited. From Fig.\ \ref{fig_NiO}A we note that the incident X-ray flux equals 50 photons/(pixel$\times$s) (i.e. 200 photons/(eV$\times$s)). The spectrometer is operated in counting mode, therefore the spectral quality will be directly correlated with the statistical error resulting from measured count rates. Consequently, for an acquisition time of 1\,h the error of the measurement at the level of 0.12\% for $I_0$, and 0.33\% for $I_1$ is achieved. The latter numbers may be used as estimate for detection limits of the setup.

\subsection{Effect of background on the spectral intensities} \label{background}

\begin{figure}
 \includegraphics[width=\columnwidth]{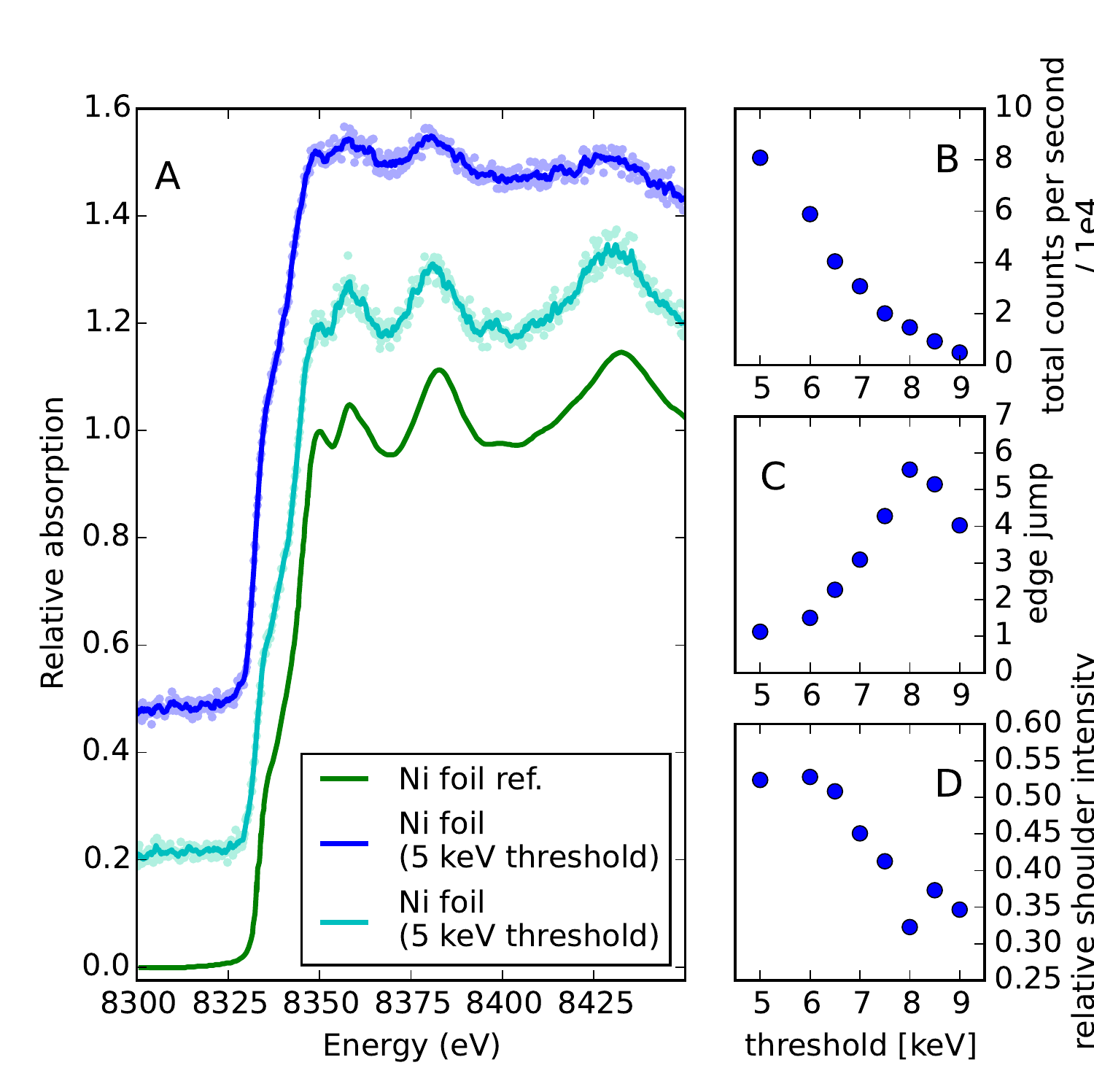}
\caption{(Color online) \textbf{A:} K-edge XANES of a Ni metal foil recorded for 10\,min on the laboratory XAS spectrometer with 5\,keV and 8\,keV thresholds (light blue and cyan dots, respectively, plotted also with blue and cyan lines using Savitzky–Golay filtering). The X-ray tube was operated with 10\,kV voltage and 40\,mA current. The $I_0$ background was measured for 10\,min. The reference spectrum is from a transmission mode scan using a Si(111) double crystal monochromator on a 3.5\,GeV\,/\,300\,mA beam produced by a third generation synchrotron \cite{Pan2012}. The laboratory spectra are shifted vertically for clarity. \textbf{B:} Total counts for the raw transmission spectrum for the Ni foil as a function of detector threshold. \textbf{C:} Change of the relative jump of the absorption edge in the absorption spectra with the detector threshold in arbitrary units (see text for details). \textbf{D:} Intensity of the Ni foil edge shoulder (at about 8336\,eV) compared to the total edge jump with different threshold settings.} \label{fig_Nifoil}
\end{figure}

In order to investigate the effect of the background, first we recorded a set of quick XANES spectra in the laboratory on a Ni-foil, for a short (10\,min) acquisition time, and at two different detector threshold settings (5 and 8\,keV). These are shown in Figure\ \ref{fig_Nifoil}, in raw and smoothed form, and they are compared to synchrotron data. While the absorption jump is very clearly visible in all spectra, at the lower threshold the fine structure appears smeared out, as it was also shown in Fig.\ \ref{fig_NiO}B for the case of NiO. The reference and the higher threshold spectra shows a reasonable match, and all main features are visible in the laboratory spectrum, but a careful comparison reveals that it has an extra broadening due to a slightly decreased resolution of 2\,eV. Although the non-monochromatic nature of our source does not allow the exact determination of the spectral resolution, this decrease may only come from the relatively large source size, which in our case is 0.4\,mm in the dispersive direction. Both the analyzer crystal and the applied pixel detector have been used at synchrotrons where they were proved to be able to provide an energy resolution well below 1\,eV \cite{szlachetko2012rsi}. However, the present 2\,eV resolution is more than enough to reveal most of the features of the XANES spectra, and the flux rate enables one to record a spectrum with sufficient quality in a few tens of minutes.

The effect of threshold on the absorption spectra is followed in the right panels of Fig.\ \ref{fig_Nifoil} with three characteristic parameters. As expected, the total number of photons recorded with the detector decreases monotonically with the increase of the threshold (Fig.\ \ref{fig_Nifoil}B). However, the relative absorption jump in the absorption spectrum, calculated as the difference between the absorbance below and above the edge and normalized to the absorbance below the edge, increases with higher thresholds until about 8\,keV (Fig.\ \ref{fig_Nifoil}C). Finally, a characteristic and well comparable value is the relative intensities of different spectral features, for example, the height of the low-energy ($\approx$\,8335 eV) shoulder to the edge in the normalized Ni foil absorption edge, which is around 0.35 in the reference spectra. This parameter is substantially deformed with lower thresholds, i.e. with considerably higher background flux both from other harmonics from the analyzer crystal and scattering fluorescence, as shown in Fig.\ \ref{fig_Nifoil}D.

\begin{figure}
\includegraphics[width=\columnwidth]{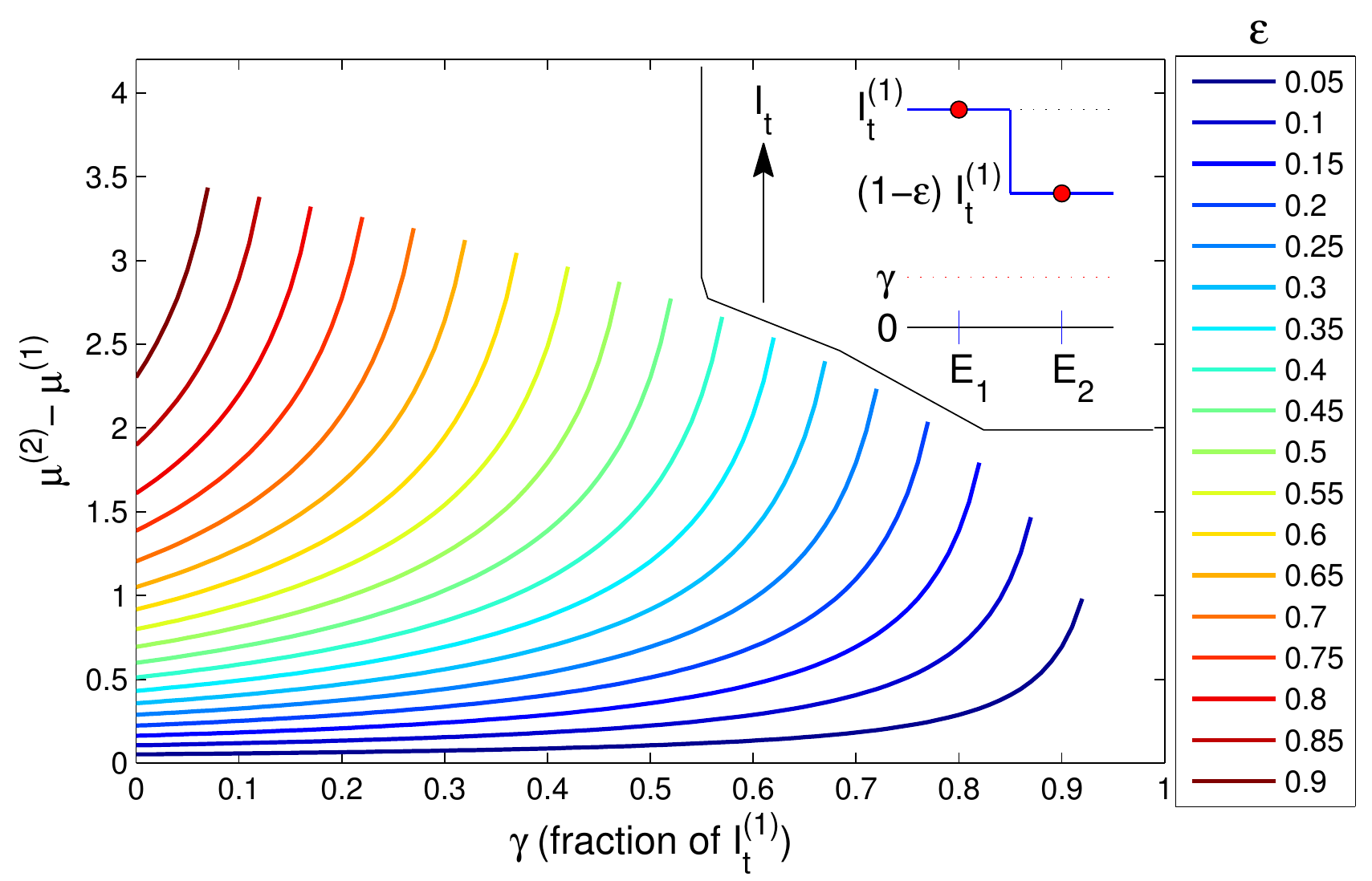}
\caption{(Color online) $\Delta \mu$ values for a pre- (1) and post-edge (2) point at different absorption values and background levels. At a certain absorption coefficient $\varepsilon$, the observed $\Delta \mu$ value is the reading at $\gamma=0$, while the real value is the reading at the actual $\gamma$ value. The inset displays a scheme with transmitted intensities at the two points as well as the edge.} \label{fig_bg}
\end{figure}

With a simple model we can examine how the observed $\mu$ (absorption coefficient) values vary with the background. For this, we select a pre-edge ($E_1$) and a post edge ($E_2$) energy point in a hypothetical XAS spectrum, expressing the intensities scaled to the transmitted intensity observed at the pre-edge, ${I}_{t}^{(1)}$. Thus the intensity drop at $E_2$ is ${I}_{t}^{(2)}-{I}_{t}^{(1)} = \varepsilon {I}_{t}^{(1)}$. The background at the two points can be approximated to be identical, denoted as $\gamma$, where $0 \geq \gamma \geq (1-\varepsilon) {I}_{t}^{(1)}$. This is illustrated in the inset of Fig.\ \ref{fig_bg}.
The absorption difference is described by the relation that follows:

\begin{equation}
\left( {\mu}^{(2)}-{\mu}^{(1)} \right) x =\ln \frac{{I}_{t}^{(1)}-\gamma}{(1-\epsilon) {I}_{t}^{(1)} - \gamma} 
\label{eq_bg}
\end{equation}

The effect of the background, modeled by Eq.\ \ref{eq_bg} is illustrated in Fig.\ \ref{fig_bg} at different absorption jumps ($\varepsilon$) and different background levels ($\gamma$). As it is seen from the figure, the distorsion effect of the background can be very large.

Next, we examine in a more detailed manner how spectra measured with background can be treated and compared to references. In case of harmonic contamination from the analyzer, the measured intensity without (${I}_{0}^{m}$) and with (${I}_{t}^{m}$) the sample can be approximated with the following terms:
\begin{eqnarray}
{I}_{0}^{m} = I_0 + {I}_{0}^{harm} + B_0 \approx I_0 (1+\alpha ) +B_0; \\
{I}_{t}^{m} = I_0 e^{-\mu x}+ \beta {I}_{0}^{harm} + B_t
\end{eqnarray}

where ${I}_{0}$ is the intensity of the radiation from the desired reflection, while ${I}_{0}^{harm}$ is the background from other harmonics (approximated as $\alpha {I}_{0}$), while $B_0$ and $B_t$ is the background from fluorescence and scattering. Using the above relations, the absorption can be expressed as

\begin{equation}
\begin{aligned}
\mu x = \ln \left( \frac{1}{1+\alpha} \frac{{I}_{0}^{m}-B_0} {{I}_{t}^{m} - \left({I}_{0}^{m}-B_0 \right)\alpha\beta/(1+\alpha) - B_t} \right) 
\end{aligned}
\label{eq_mux}
\end{equation}

In order to arrive to a model that can be directly compared with reference spectra, we need to introduce some further approximations, normalize the data to the unit edge jump, after flattening the pre-edge and the post-edge with the usual linear and quadratic approximations, respectively. With these, we arrive to what corresponds to the XAS $\chi(E)$, as seen in Eq.\ \ref{eq_fitmodel}. 

\begin{equation}
\begin{aligned}
\chi(E) & {} =  P_1 \left[ \ln \frac{{I}_{0}^{m}-P_2} {{I}_{t}^{m} - \left({I}_{0}^{m}-P_2 \right)P_3 - P_4} - P_5 E -P_6 \right. \\
	& \left. - H\left(E-P_7\right) \left( {P_8\left(E-P_7\right)}^{2} + P_9 \left(E-P_7\right) +P_{10} \right) \vphantom{\frac{{I}_{0}^{m}} {{I}_{t}^{m}}}\right]
\end{aligned}
\label{eq_fitmodel}
\end{equation}

The above equation can be fitted to the data to separate the background contributions and recover the true signal, with the $P_{\mathrm i}$ as fit parameters. ($P_1$ is a scale factor to produce unit edge jump, $P_2$ and $P_4$ are $B_0$ and $B_t$ approximated with constants, respectively, $P_3$ corresponds to the $\alpha\beta/(1+\alpha)$ factor scaling the intensity through the wrong harmonics, $P_5$ and $P_6$ are for the linear pre-edge, and $P_7$ is the $E_0$, and $P_8, P_9$ and $P_{10}$ flattens the post-edge, which is achieved with the heavyside step function $H$. (Rigorously, $P_2$, and $P_4$ are vectors corresponding to the length of the spectra with the identical $P_2$ or $P_4$ value for each components.)

\begin{figure}
\includegraphics[width=\columnwidth]{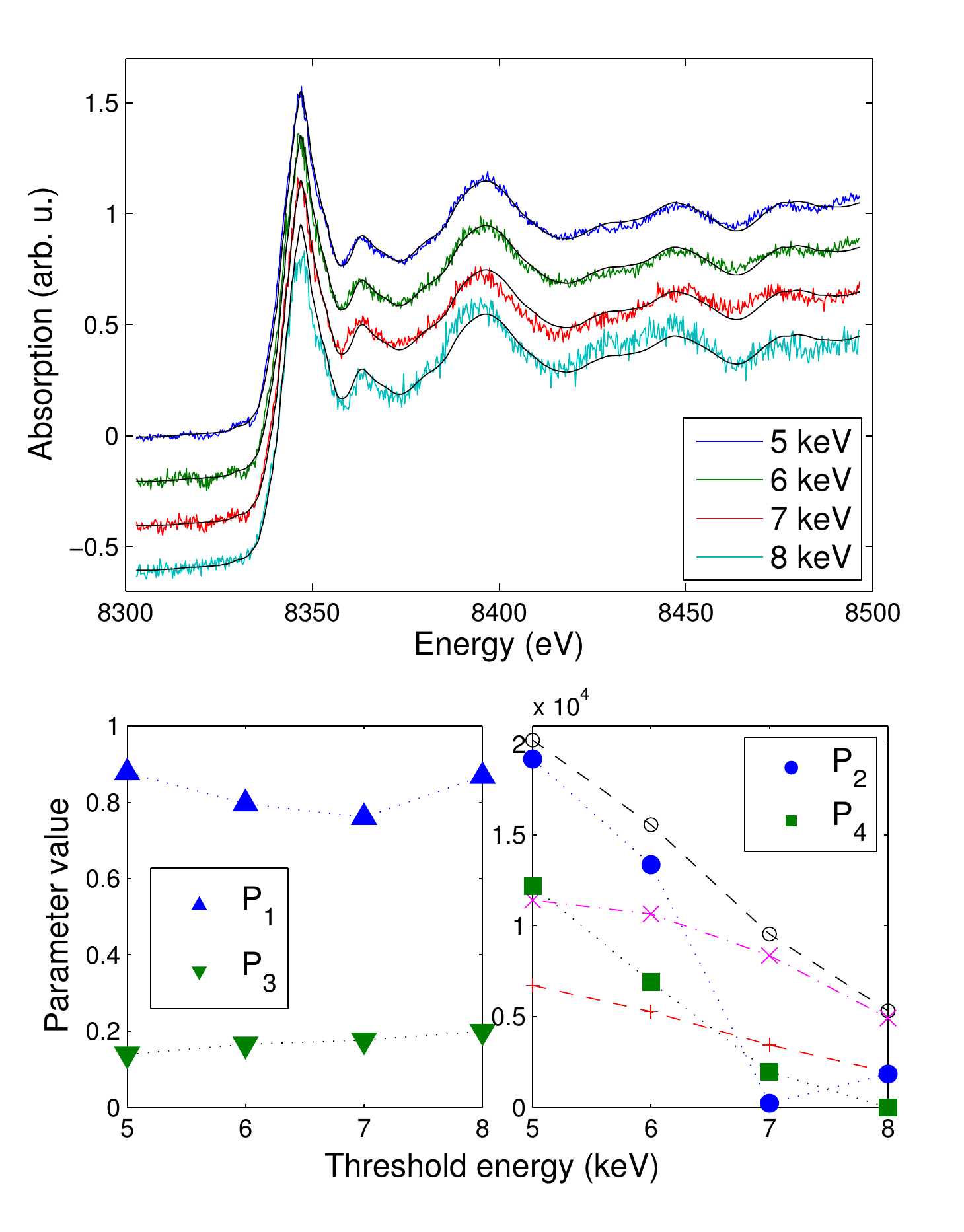}
\caption{(Color online) (top) NiO spectra taken for 10 minutes at different detector thresholds between 5 keV and 8 keV, fitted with Eq.\ \ref{eq_fitmodel} to the reference spectra (shown in black). The bottom shows how the 4 most relevant parameter varies with the threshold. Also shown in the bottom right panel how the the mean $I^m_0/5$ (o), the mean $I^m_t/5$ (+) and the (mean$(I^m_0)-P_2)P_3$ (x) varies with the threshold.} \label{fig_fits}
\end{figure}

Using the model in Eq.\ \ref{eq_fitmodel} to fit NiO spectra taken with different detector thresholds, the absorption spectrum can be reconstructed to reasonable accuracy, as shown in Fig.\ \ref{fig_fits}. The background parameters $P_2$ and $P_4$ show a decrease with the increase in the threshold, and while $P_3$ seems to slightly increase, its product with $({I}_{0}^{m}-P_2)$, which corresponds to the harmonic contamination, drops as the latter term also decreases rapidly. Further insights are blocked by the too many parameters in the model, which give rise to large correlations between physically formally independent parameters. 

\subsection{Chemical sensitivity of K-edge XAS, at synchrotrons and in the lab} 

One of the most promising applications of a laboratory XAS spectrometer is the utilization of the high chemical sensitivity of the absorption edge of the selected metal atom. It is well known that the absorption edge position shifts towards higher energies with higher oxidation state of the investigated element due to the increase of the effective nuclear charge, even if a strict relationship between edge position and oxidation state can only be established for compounds with similar coordination sphere and approximately isotropic variations.\cite{glatzel2009xafs14} During the past decades a complete set of absorption edge energies has been published for different elements with different oxidation-, spin- and coordination states, and the appropriate calibration compounds are also often easily accessible. Hence this method provides a quick way to characterize the electronic structure of any atoms in novel materials with an absorption edge in the energy range of hard X-rays.

\begin{figure}
 \includegraphics[width=\columnwidth]{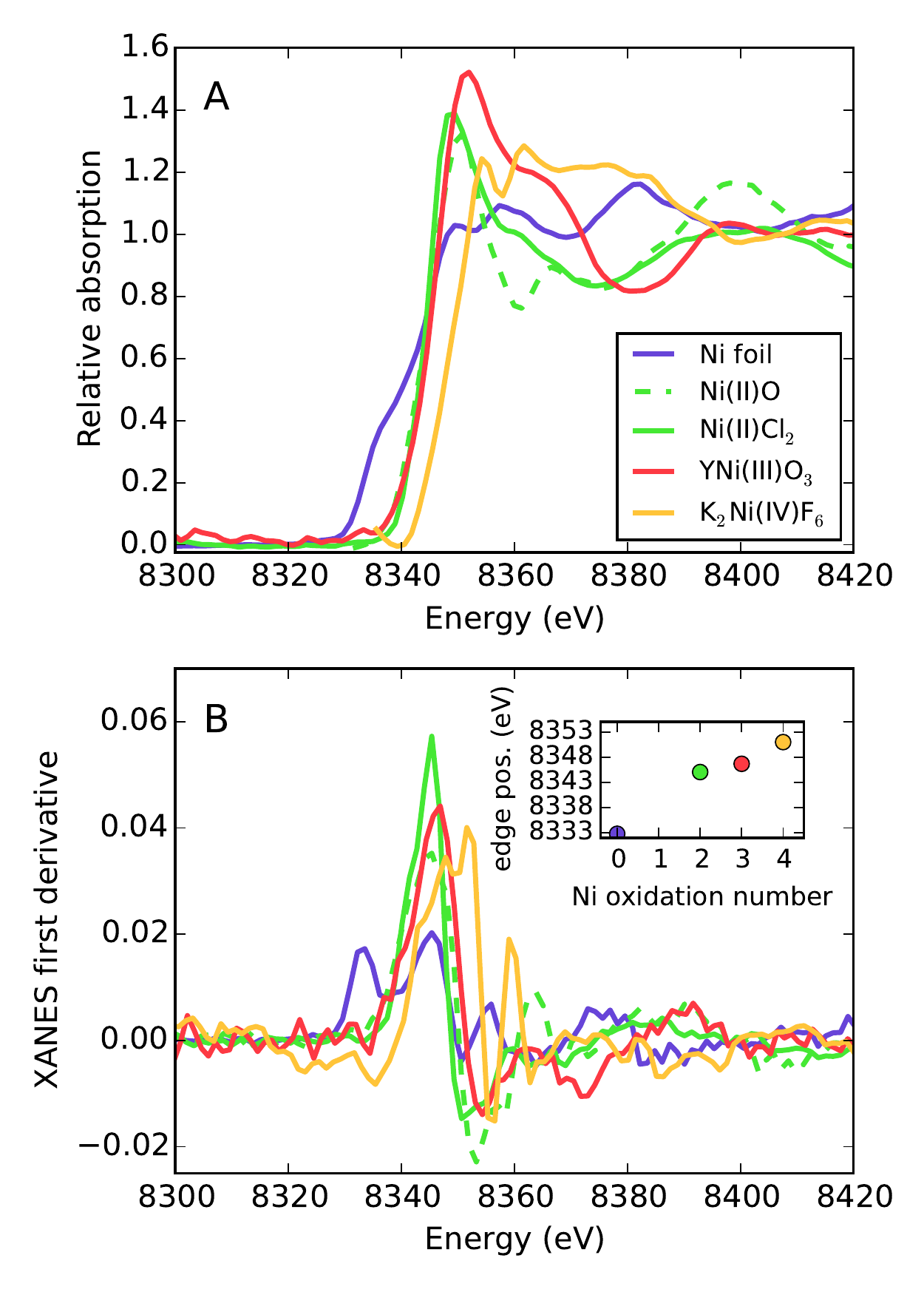}
\caption{(Color online) \textbf{A:} Laboratory XANES spectra of selected Ni compounds. Acquisition times (respective background acquisitions) were the following: 500\,min (30\,min) for NiO, 20\,min (10\,min) for \ce{NiCl2} and \ce{YNiO3} and 74\,min (140\,min) for \ce{K2NiF6} Spectra are binned with a bin size of 5 (see text for details). \textbf{B:} First derivatives of the spectra in panel A. Inset: deduced K edge positions vs. oxidation number. Error bars are smaller than the dot size.} \label{fig_Nichem}
\end{figure}

In order to test the chemical sensitivity, we measured the K edge XANES of a few distinct Ni compounds: a metallic Ni-foil, divalent NiO, divalent \ce{NiCl2}, trivalent \ce{YNiO3} and tetravalent \ce{K2NiF6}. The recorded absorption spectra (Fig.\ \ref{fig_Nichem}A) resemble well those in previous studies using synchrotron radiation \cite{Mori2015,Tirez2011,Medarde2009,Zhang2015, woolley2011,ogrady1996}. The edge positions, determined as the maximum of the first derivative, clearly show the expected increasing trend with the oxidation state of nickel (see Fig.\ \ref{fig_Nichem}B and its inset).

\begin{figure}
 \includegraphics[width=\columnwidth]{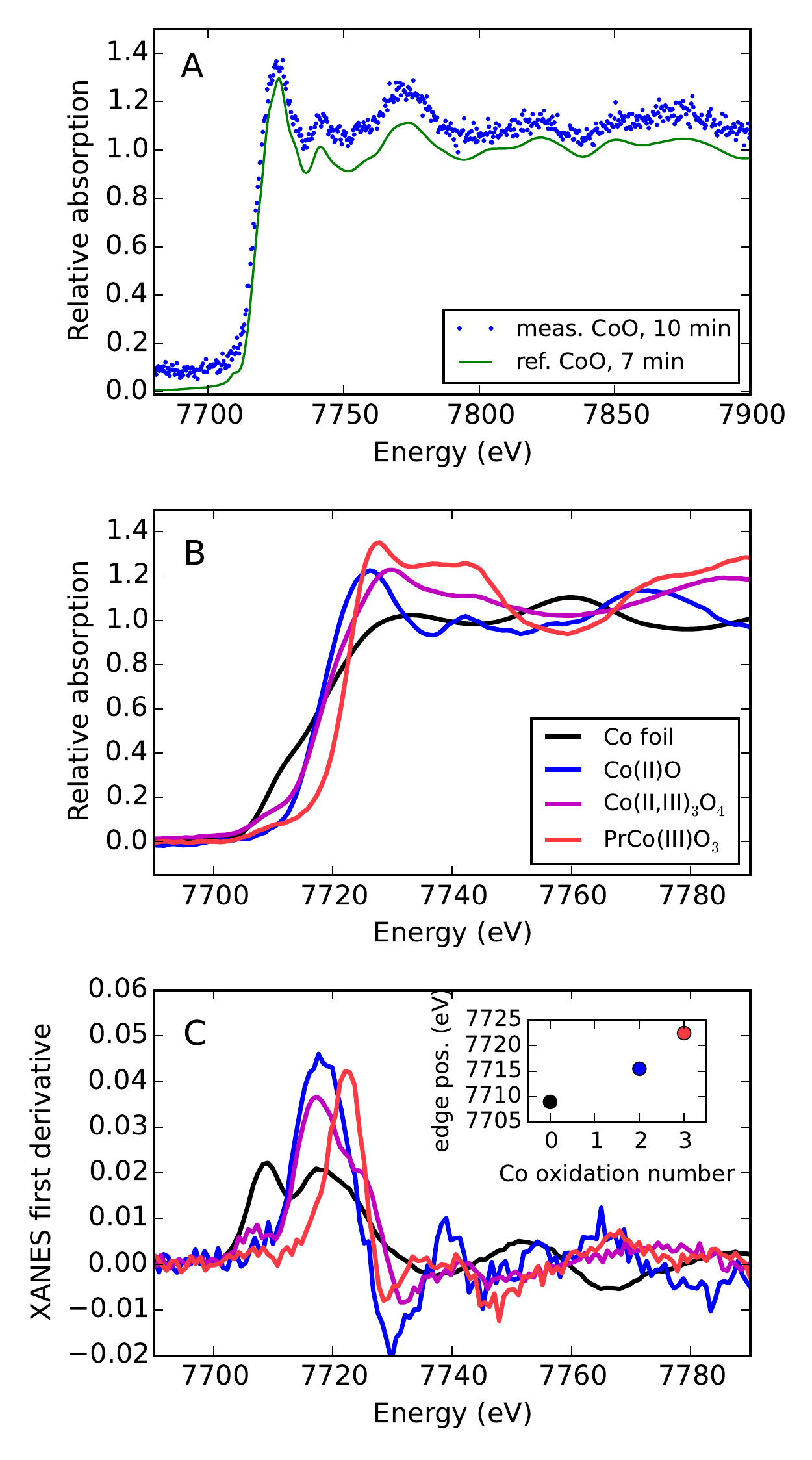}
\caption{(Color online) \textbf{A:} K-edge XANES of CoO measured with the present laboratory spectrometer for 10\,min compared to the same spectrum taken at the DORIS3 synchrotron for 7 minutes. The laboratory CoO spectrum is shifted vertically. \textbf{B:} Co K edge XANES of selected samples with different oxidation states, as denoted on the graph. Acquisition times (respective background acquisitions) were the following: 260\,min (160\,min) for Co, 10\,min (10\,min) for \ce{CoO}, 5\,min (5\,min) for \ce{Co3O4} and 210\,min (120\,min) for \ce{PrCoO3}. Spectra are binned as described in text. \textbf{C:} First derivatives of the spectra in the subplot \textbf{B}. Inset: the edge position vs. the nominal oxidation number of cobalt. Error bars are smaller than the dot size.} \label{fig_Co_chem}
\end{figure}

The flexibility of the presented setup allows us to reach other edges easily with the same setup. A linear translation of the analyzer crystal and (twice as much of) the detector to obtain 53.3$^\circ$ mean Bragg-angle at the Si(333) reflection, and the corresponding rotation of the spectrometer breadboard to bring the analyzer back in the beam makes the Co K edge accessible. In Figure\ \ref{fig_Co_chem} some selected Co K edge spectra are plotted, recorded with the laboratory spectrometer. In Fig.\ \ref{fig_Co_chem}A a comparison between the laboratory spectrum and a synchrotron scan is shown for CoO powder sample. The acquisition times (10\,min and 7\,min) are comparable, and the resolution is enough to observe all main features of the edge jump and above the edge. It is worth noting that the decreased resolution smears out the sharp pre-edge peak, but this can be addressed if a microfocus X-ray tube is used as source. Figs.\ \ref{fig_Co_chem}B and C showcase the chemical sensitivity of the present XAS spectrometer. A metallic cobalt foil, a divalent (CoO), a trivalent \ce{PrCoO3} and a mixed valence (\ce{Co3^{2+,3+}O4}) Co compound were measured and analyzed. The edge positions, deduced from the first derivative of the absorption curve, change monotonically with the oxidation state of the cobalt ion. Even the two distinct oxidation states in the mixed \ce{Co3O4} oxide can be clearly separated.

\subsection{Applications: monitoring a slow chemical transformation} 

\begin{figure}
 \includegraphics[width=\columnwidth]{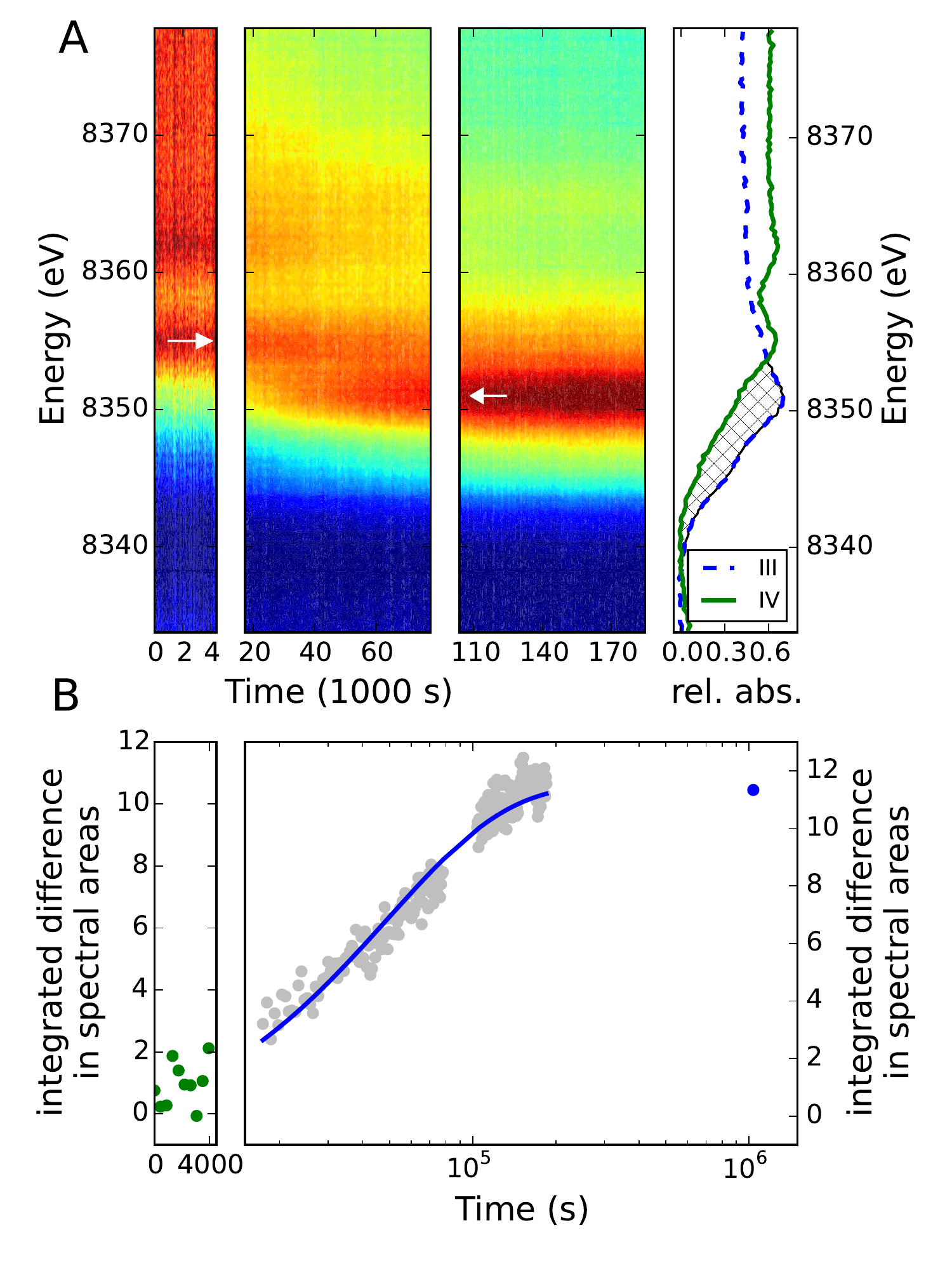}
\caption{(Color online) \textbf{A:} Left three panels: 2D mapped evolution of the Ni K edge XANES of a \ce{K2Ni(IV)F6} sample followed for 51.5\,h in three batches. Two white line maxima at around 8356\,eV and 8351\,eV (indicated by white arrows) can be clearly separated, corresponding to the tetravalent and trivalent states of nickel, respectively. The panels share the same color scale. The right panel shows the XANES spectrum of the original and final product recorded after 12\,days, denoted after the valence state of Ni as IV and III, respectively. \textbf{B:} Integrated differential signal intensity between 8338\,eV and 8354\,eV energies (shown as gray patterned area in the right panel of subplot A), corresponding to the change from the Ni(IV) to Ni(III) state. The trivalent state is reached after ca. 41\,h. The first order kinetic fit is shown as a blue line on the gray dots.} \label{fig_NiIV}
\end{figure}

A typical case which cannot be performed at a synchrotron is the monitoring of slow chemical reactions. To demonstrate this, our selected example is the hydrolysis of \ce{K2Ni(IV)F6} with humidity in the air. In this compound the unstable tetravalent state of the nickel ion reduces to trivalent in the presence of water. Depending on how efficiently the powder is sealed from air, this transformation can last for days. Figure\ \ref{fig_NiIV} shows the evolution of the Ni XANES as a function of time. The spectra were recorded in four time slices: a short acquisition for the first 73 minutes with 1 min steps, a longer one after 3.5\,h with 10 min steps for ca.\ 17 hours, a third batch after 29\,h for ca.\ 23 hours, and a last set after 12\,days to check the final state of the sample. The observed spectra tracks clearly the reduction of the Ni(IV) ions. The absorption edge maximum corresponding to the tetravalent state (at around 8356\,eV) fades away in the first hours giving rise to a second absorption edge peaking at around 8351\,eV. This latter stabilizes at the end of the second measurement series and remains the dominant component even days after, as shown in the right panel of Fig.\ \ref{fig_NiIV}A. In order to get a quantitative description of the reaction kinetics, in Fig.\ \ref{fig_NiIV}B the difference in the two spectral areas integrated between 8338\,eV and 8354\,eV is plotted, these energies correspond to the two isosbestic points before and after the absorption edge of the trivalent component compared to the one of the tetravalent one. This area can help to easily follow the concentration of the evolving species. The graph depicts an apparent simple exponential kinetics with a first order rate constant of $2.03 \cdot 10^{-5}$\,1/s. This demonstrates that with a careful control of the experimental conditions, such measurements can be exploited to study kinetics of slow transport or transformation processes.

\subsection{Laboratory XES with the von H\'amos setup} 

\begin{figure}
 \includegraphics[width=\columnwidth]{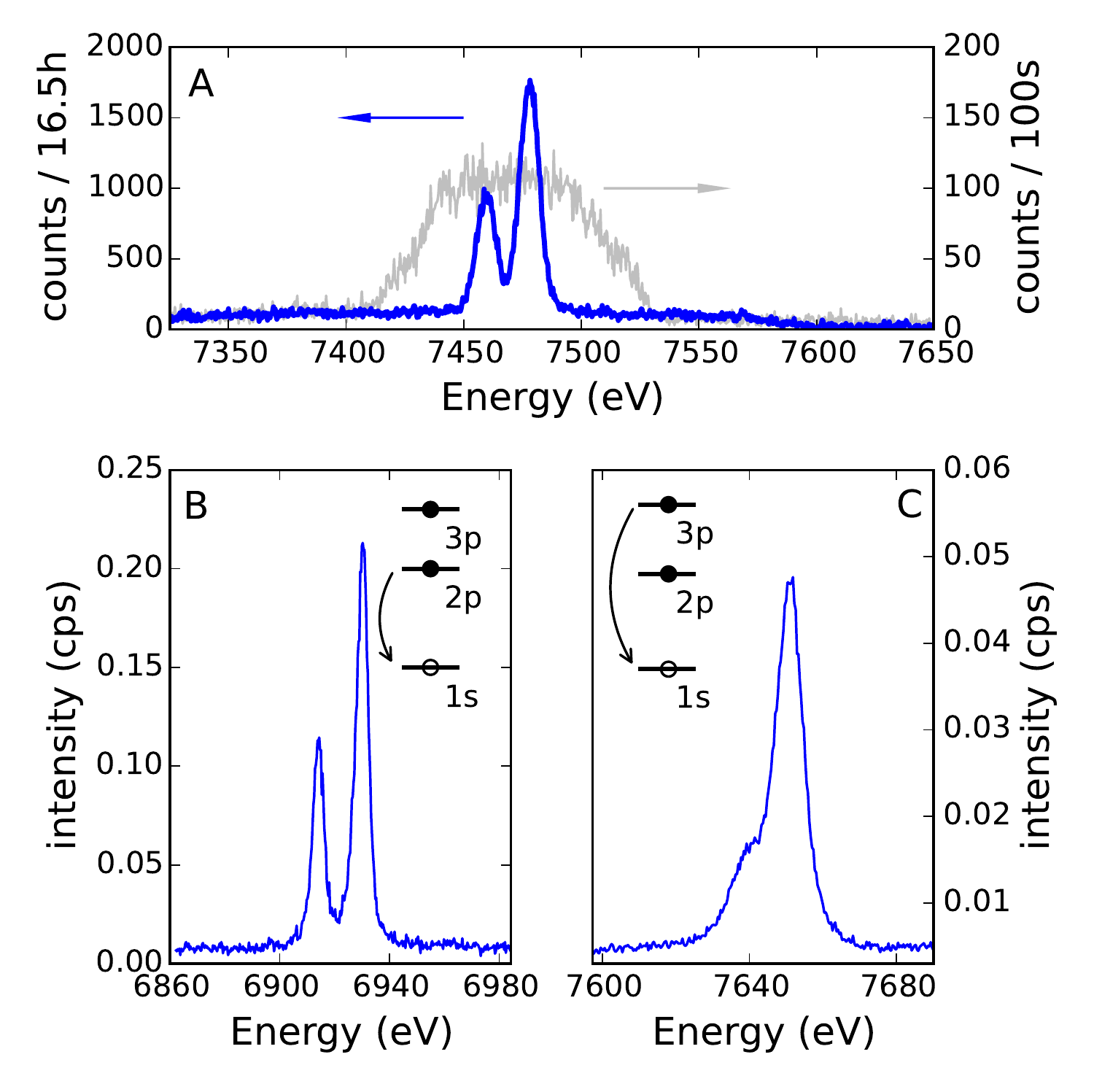}
\caption{(Color online) \textbf{A:} K$\alpha$ X-ray emission spectra of a Ni foil with and without slitting the sample (blue and gray curves, respectively). Panels \textbf{B} and \textbf{C} show the K$\alpha$ and K$\beta$ spectra of a CoO sample. The CoO K$\alpha$ was recorded for 70 min, while the K$\beta$ was taken in 19\,h. The insets represent the corresponding electronic transitions of the Co K$\alpha$ and K$\beta$ fluorescence.} \label{fig_xes}
\end{figure}

XAS and XES spectroscopies are complementary techniques in terms of information provided. The laboratory X-ray spectrometer can be quickly realigned to be able to record XES spectra by rotating and shifting its breadboard to bring the sample in the source position, as described in Section \ref{sec:xes}. Figure\ \ref{fig_xes}A shows two K$\alpha$ spectra of a Ni-foil recorded on our setup. The incident radiation on the sample was generated in the X-ray tube with 10\,keV and 60\,mA settings. The gray line showcases the spectra from a larger Ni-foil, with an estimated apparent source size of 2\,mm in the projection to the analyzer crystal. The resolution in this case is clearly insufficient to resolve the K$\alpha_1$ and K$\alpha_2$ lines. Although slitting the sample emission results in a decrease in total fluorescent photon intensity, the resolution can be drastically increased up to the limit set by the analyzer, e.g. comparable to standard $3^{\rm rd}$ generation synchrotron XES spectrometers. The blue curve in Fig.\ \ref{fig_xes}A represents the latter case. The slit applied was in the order of 0.1\,mm in the projection onto the crystal.

Besides the most intense K$\alpha$ lines, the spectrometer is well capable of recording decent quality K$\beta$ spectra, as well. Figs.\ \ref{fig_xes}B and C show both K$\alpha$ and K$\beta$ emission spectra of CoO, acquired for 70\,min and 19\,h, respectively. Although the count rate with the applied slit and flux is rather limited, the K$\beta_{1,3}$ line at 7649\,eV and its K$\beta'$ shoulder around 7640\,eV, which is one of the most sensitive features to identify the spin state in transition metals \cite{vanko2006jpcb}, can be recorded with satisfactory signal-to-noise ratio. 

\section{Summary and Outlook}

We have presented a laboratory scale, cost effective high resolution X-ray spectrometer based on von H\'amos single crystal analyzers, capable of recording both XAS and XES spectra with a standard X-ray tube in acceptable time periods. The spectrometer consists of commercially available components, and provides a flexible, quickly adjustable setup. One of the key advantages of this spectrometer is that it does not need any scanning components. A fixed and stable setup can be operated even for long acquisition times. Moreover, with a 25\,cm bending radius crystal the intensity loss in the spectrometer due to air absorption can be tolerated for many edges, which makes the setup simpler and cheaper compared to the previously reported ones \cite{Seidler2014}. The single crystal analyzer and the high resolution position sensitive detector ensures a resolution comparable to X-ray spectrometers at $3^{\rm rd}$ generation synchrotrons.

While the present paper demonstrates the potential of this spectrometer, further extensions can help to improve its capabilities. A major shortcoming of the present setup is the relatively large source size, and the resulting decrease in spectral resolution. This can be resolved by a using microfocus X-ray source, which provides an X-ray beam focused to a few tens of micrometers, sufficient for high resolution XAS and XES experiments. In addition, using a microscopic X-ray beam, the chemical inhomogenities will no longer modulate the spectra, enabling one to combine several XAS regions with different energy ranges to cover the whole EXAFS region. Finally, while in high background (low S/N) cases, such as measuring pre-edges or valence-to-core XES, Johann-spectrometers seem more advantageous, it is to be examined how the von H\'amos setup will perform with the higher brilliance microfocus sources.

A simpler version of the present setup could be based on a set of interchangable breadboards with optimized, fixed crystal- and detector positions, where one can easily switch between XAS and XES setups or between different elements, making the instrument flexible, more economic and user friendly. This spectrometer has the potential to allow the wider scientific community to access the unique information provided by hard X-ray spectroscopies while simultaneously allowing overbooked synchrotron facilities to focus on challenging experiments to which they are better suited.

\begin{acknowledgments}
This project was supported by the European Research Council via contract ERC-StG-259709 (X-cited!), and the 'Lend{\"u}let' (Momentum) Program of the Hungarian Academy of Sciences (LP2013-59). ZN acknowledges support from the Bolyai Fellowship of the Hungarian Academy of Sciences. We thank Csilla Bogd\'an, Francesco Raiola and G\'abor Rozip\'al for assistance at different phases of this project, and Chris Milne and Jesse Weil for their comments on the manuscript.
\end{acknowledgments}

\bibliography{refs_labvH.bib}

\begin{thebibliography}{32}%
\makeatletter
\providecommand \@ifxundefined [1]{%
 \@ifx{#1\undefined}
}%
\providecommand \@ifnum [1]{%
 \ifnum #1\expandafter \@firstoftwo
 \else \expandafter \@secondoftwo
 \fi
}%
\providecommand \@ifx [1]{%
 \ifx #1\expandafter \@firstoftwo
 \else \expandafter \@secondoftwo
 \fi
}%
\providecommand \natexlab [1]{#1}%
\providecommand \enquote  [1]{``#1''}%
\providecommand \bibnamefont  [1]{#1}%
\providecommand \bibfnamefont [1]{#1}%
\providecommand \citenamefont [1]{#1}%
\providecommand \href@noop [0]{\@secondoftwo}%
\providecommand \href [0]{\begingroup \@sanitize@url \@href}%
\providecommand \@href[1]{\@@startlink{#1}\@@href}%
\providecommand \@@href[1]{\endgroup#1\@@endlink}%
\providecommand \@sanitize@url [0]{\catcode `\\12\catcode `\$12\catcode
  `\&12\catcode `\#12\catcode `\^12\catcode `\_12\catcode `\%12\relax}%
\providecommand \@@startlink[1]{}%
\providecommand \@@endlink[0]{}%
\providecommand \url  [0]{\begingroup\@sanitize@url \@url }%
\providecommand \@url [1]{\endgroup\@href {#1}{\urlprefix }}%
\providecommand \urlprefix  [0]{URL }%
\providecommand \Eprint [0]{\href }%
\providecommand \doibase [0]{http://dx.doi.org/}%
\providecommand \selectlanguage [0]{\@gobble}%
\providecommand \bibinfo  [0]{\@secondoftwo}%
\providecommand \bibfield  [0]{\@secondoftwo}%
\providecommand \translation [1]{[#1]}%
\providecommand \BibitemOpen [0]{}%
\providecommand \bibitemStop [0]{}%
\providecommand \bibitemNoStop [0]{.\EOS\space}%
\providecommand \EOS [0]{\spacefactor3000\relax}%
\providecommand \BibitemShut  [1]{\csname bibitem#1\endcsname}%
\let\auto@bib@innerbib\@empty
\bibitem [{\citenamefont {Bunker}(2010)}]{Bunker2010}%
  \BibitemOpen
  \bibfield  {author} {\bibinfo {author} {\bibfnamefont {G.}~\bibnamefont
  {Bunker}},\ }\href@noop {} {\emph {\bibinfo {title} {Introduction to XAFS A
  Practical Guide to X-ray Absorption Fine Structure Spectroscopy}}},\ edited
  by\ \bibinfo {editor} {\bibfnamefont {G.}~\bibnamefont {Bunker}}\ (\bibinfo
  {publisher} {Cambridge University Press},\ \bibinfo {year}
  {2010})\BibitemShut {NoStop}%
\bibitem [{\citenamefont {Koningsberger}\ and\ \citenamefont
  {Prins}(1987)}]{Koningsberger1987}%
  \BibitemOpen
  \bibfield  {author} {\bibinfo {author} {\bibfnamefont {D.}~\bibnamefont
  {Koningsberger}}\ and\ \bibinfo {author} {\bibfnamefont {R.}~\bibnamefont
  {Prins}},\ }\href@noop {} {\emph {\bibinfo {title} {X-Ray Absorption
  Principles, Applications, Techniques of EXAFS, SEXAFS and XANES Chemical
  Analysis: A Series of Monographs on Analytical Chemistry and Its
  Applications}}},\ edited by\ \bibinfo {editor} {\bibfnamefont
  {D.}~\bibnamefont {Koningsberger}}\ and\ \bibinfo {editor} {\bibfnamefont
  {R.}~\bibnamefont {Prins}}\ (\bibinfo  {publisher} {Wiley \& Sons Ltd},\
  \bibinfo {year} {1987})\BibitemShut {NoStop}%
\bibitem [{\citenamefont {Stern}\ and\ \citenamefont
  {Heald}(1983)}]{Stern1983}%
  \BibitemOpen
  \bibfield  {author} {\bibinfo {author} {\bibfnamefont {E.~A.}\ \bibnamefont
  {Stern}}\ and\ \bibinfo {author} {\bibfnamefont {S.~M.}\ \bibnamefont
  {Heald}},\ }\href@noop {} {\emph {\bibinfo {title} {Handbook of Synchrotron
  Radiation: Basic Principles and Applications of EXAFS}}},\ edited by\
  \bibinfo {editor} {\bibfnamefont {E.~E.}\ \bibnamefont {Koch}},\ Vol.\
  \bibinfo {volume} {Chapter 10}\ (\bibinfo  {publisher} {North-Holland},\
  \bibinfo {year} {1983})\ pp.\ \bibinfo {pages} {pp 995--1014}\BibitemShut
  {NoStop}%
\bibitem [{\citenamefont {Glatzel}\ and\ \citenamefont
  {Bergmann}(2005)}]{glatzel2005ccr}%
  \BibitemOpen
  \bibfield  {author} {\bibinfo {author} {\bibfnamefont {P.}~\bibnamefont
  {Glatzel}}\ and\ \bibinfo {author} {\bibfnamefont {U.}~\bibnamefont
  {Bergmann}},\ }\href {\doibase 10.1016/j.ccr.2004.04.011} {\bibfield
  {journal} {\bibinfo  {journal} {Coord. Chem. Rev.}\ }\textbf {\bibinfo
  {volume} {249}},\ \bibinfo {pages} {65} (\bibinfo {year} {2005})}\BibitemShut
  {NoStop}%
\bibitem [{\citenamefont {de~Groot}(2001)}]{deGroot2001}%
  \BibitemOpen
  \bibfield  {author} {\bibinfo {author} {\bibfnamefont {F.~M.~F.}\
  \bibnamefont {de~Groot}},\ }\href@noop {} {\bibfield  {journal} {\bibinfo
  {journal} {Chem. Rev.}\ }\textbf {\bibinfo {volume} {101}},\ \bibinfo {pages}
  {1779} (\bibinfo {year} {2001})}\BibitemShut {NoStop}%
\bibitem [{\citenamefont {Torchio}, \citenamefont {Mathon},\ and\ \citenamefont
  {Pascarelli}(2014)}]{torchio2014}%
  \BibitemOpen
  \bibfield  {author} {\bibinfo {author} {\bibfnamefont {R.}~\bibnamefont
  {Torchio}}, \bibinfo {author} {\bibfnamefont {O.}~\bibnamefont {Mathon}}, \
  and\ \bibinfo {author} {\bibfnamefont {S.}~\bibnamefont {Pascarelli}},\
  }\href {\doibase 10.1016/j.ccr.2014.02.024} {\bibfield  {journal} {\bibinfo
  {journal} {Coordination Chemistry Reviews}\ }\textbf {\bibinfo {volume}
  {277--278}},\ \bibinfo {pages} {80 } (\bibinfo {year} {2014})}\BibitemShut
  {NoStop}%
\bibitem [{\citenamefont {Rueff}\ and\ \citenamefont
  {Shukla}(2010)}]{rueff2010}%
  \BibitemOpen
  \bibfield  {author} {\bibinfo {author} {\bibfnamefont {J.-P.}\ \bibnamefont
  {Rueff}}\ and\ \bibinfo {author} {\bibfnamefont {A.}~\bibnamefont {Shukla}},\
  }\href {\doibase 10.1103/RevModPhys.82.847} {\bibfield  {journal} {\bibinfo
  {journal} {Rev. Mod. Phys.}\ }\textbf {\bibinfo {volume} {82}},\ \bibinfo
  {pages} {847} (\bibinfo {year} {2010})}\BibitemShut {NoStop}%
\bibitem [{\citenamefont {Rovezzi}\ and\ \citenamefont
  {Glatzel}(2014)}]{rovezzi2014}%
  \BibitemOpen
  \bibfield  {author} {\bibinfo {author} {\bibfnamefont {M.}~\bibnamefont
  {Rovezzi}}\ and\ \bibinfo {author} {\bibfnamefont {P.}~\bibnamefont
  {Glatzel}},\ }\href {http://stacks.iop.org/0268-1242/29/i=2/a=023002}
  {\bibfield  {journal} {\bibinfo  {journal} {Semiconductor Science and
  Technology}\ }\textbf {\bibinfo {volume} {29}},\ \bibinfo {pages} {023002}
  (\bibinfo {year} {2014})}\BibitemShut {NoStop}%
\bibitem [{\citenamefont {Seidler}\ \emph {et~al.}(2014)\citenamefont
  {Seidler}, \citenamefont {Mortensen}, \citenamefont {Remesnik}, \citenamefont
  {Pacold}, \citenamefont {Ball}, \citenamefont {Barry}, \citenamefont
  {Styczinski},\ and\ \citenamefont {Hoidn}}]{Seidler2014}%
  \BibitemOpen
  \bibfield  {author} {\bibinfo {author} {\bibfnamefont {G.~T.}\ \bibnamefont
  {Seidler}}, \bibinfo {author} {\bibfnamefont {D.~R.}\ \bibnamefont
  {Mortensen}}, \bibinfo {author} {\bibfnamefont {A.~J.}\ \bibnamefont
  {Remesnik}}, \bibinfo {author} {\bibfnamefont {J.~I.}\ \bibnamefont
  {Pacold}}, \bibinfo {author} {\bibfnamefont {N.~A.}\ \bibnamefont {Ball}},
  \bibinfo {author} {\bibfnamefont {N.}~\bibnamefont {Barry}}, \bibinfo
  {author} {\bibfnamefont {M.}~\bibnamefont {Styczinski}}, \ and\ \bibinfo
  {author} {\bibfnamefont {O.~R.}\ \bibnamefont {Hoidn}},\ }\href {\doibase
  10.1063/1.4901599} {\bibfield  {journal} {\bibinfo  {journal} {Review of
  Scientific Instruments}\ }\textbf {\bibinfo {volume} {85}},\ \bibinfo {eid}
  {113906} (\bibinfo {year} {2014})}\BibitemShut {NoStop}%
\bibitem [{\citenamefont {{Mortensen}}\ \emph {et~al.}(2015)\citenamefont
  {{Mortensen}}, \citenamefont {{Seidler}}, \citenamefont {{Ditter}},\ and\
  \citenamefont {{Glatzel}}}]{MortensenarXiv2015}%
  \BibitemOpen
  \bibfield  {author} {\bibinfo {author} {\bibfnamefont {D.~R.}\ \bibnamefont
  {{Mortensen}}}, \bibinfo {author} {\bibfnamefont {G.~T.}\ \bibnamefont
  {{Seidler}}}, \bibinfo {author} {\bibfnamefont {A.~S.}\ \bibnamefont
  {{Ditter}}}, \ and\ \bibinfo {author} {\bibfnamefont {P.}~\bibnamefont
  {{Glatzel}}},\ }\href@noop {} {\bibfield  {journal} {\bibinfo  {journal}
  {ArXiv e-prints}\ } (\bibinfo {year} {2015})},\ \Eprint
  {http://arxiv.org/abs/1509.05711} {arXiv:1509.05711 [cond-mat.mtrl-sci]}
  \BibitemShut {NoStop}%
\bibitem [{\citenamefont {{Seidler}}\ \emph {et~al.}(2015)\citenamefont
  {{Seidler}}, \citenamefont {{Mortensen}}, \citenamefont {{Ditter}},
  \citenamefont {{Ball}},\ and\ \citenamefont {{Remesnik}}}]{SeidlerarXiv2015}%
  \BibitemOpen
  \bibfield  {author} {\bibinfo {author} {\bibfnamefont {G.~T.}\ \bibnamefont
  {{Seidler}}}, \bibinfo {author} {\bibfnamefont {D.~R.}\ \bibnamefont
  {{Mortensen}}}, \bibinfo {author} {\bibfnamefont {A.~S.}\ \bibnamefont
  {{Ditter}}}, \bibinfo {author} {\bibfnamefont {N.~A.}\ \bibnamefont
  {{Ball}}}, \ and\ \bibinfo {author} {\bibfnamefont {A.~J.}\ \bibnamefont
  {{Remesnik}}},\ }\href@noop {} {\bibfield  {journal} {\bibinfo  {journal}
  {ArXiv e-prints}\ } (\bibinfo {year} {2015})},\ \Eprint
  {http://arxiv.org/abs/1509.05708} {arXiv:1509.05708 [physics.ins-det]}
  \BibitemShut {NoStop}%
\bibitem [{\citenamefont {Uhlig}\ \emph {et~al.}(2013)\citenamefont {Uhlig},
  \citenamefont {Fullagar}, \citenamefont {Ullom}, \citenamefont {Doriese},
  \citenamefont {Fowler}, \citenamefont {Swetz}, \citenamefont {Gador},
  \citenamefont {Canton}, \citenamefont {Kinnunen}, \citenamefont {Maasilta},
  \citenamefont {Reintsema}, \citenamefont {Bennett}, \citenamefont {Vale},
  \citenamefont {Hilton}, \citenamefont {Irwin}, \citenamefont {Schmidt},\ and\
  \citenamefont {Sundstr\"om}}]{uhlig2013}%
  \BibitemOpen
  \bibfield  {author} {\bibinfo {author} {\bibfnamefont {J.}~\bibnamefont
  {Uhlig}}, \bibinfo {author} {\bibfnamefont {W.}~\bibnamefont {Fullagar}},
  \bibinfo {author} {\bibfnamefont {J.~N.}\ \bibnamefont {Ullom}}, \bibinfo
  {author} {\bibfnamefont {W.~B.}\ \bibnamefont {Doriese}}, \bibinfo {author}
  {\bibfnamefont {J.~W.}\ \bibnamefont {Fowler}}, \bibinfo {author}
  {\bibfnamefont {D.~S.}\ \bibnamefont {Swetz}}, \bibinfo {author}
  {\bibfnamefont {N.}~\bibnamefont {Gador}}, \bibinfo {author} {\bibfnamefont
  {S.~E.}\ \bibnamefont {Canton}}, \bibinfo {author} {\bibfnamefont
  {K.}~\bibnamefont {Kinnunen}}, \bibinfo {author} {\bibfnamefont {I.~J.}\
  \bibnamefont {Maasilta}}, \bibinfo {author} {\bibfnamefont {C.~D.}\
  \bibnamefont {Reintsema}}, \bibinfo {author} {\bibfnamefont {D.~A.}\
  \bibnamefont {Bennett}}, \bibinfo {author} {\bibfnamefont {L.~R.}\
  \bibnamefont {Vale}}, \bibinfo {author} {\bibfnamefont {G.~C.}\ \bibnamefont
  {Hilton}}, \bibinfo {author} {\bibfnamefont {K.~D.}\ \bibnamefont {Irwin}},
  \bibinfo {author} {\bibfnamefont {D.~R.}\ \bibnamefont {Schmidt}}, \ and\
  \bibinfo {author} {\bibfnamefont {V.}~\bibnamefont {Sundstr\"om}},\ }\href
  {\doibase 10.1103/PhysRevLett.110.138302} {\bibfield  {journal} {\bibinfo
  {journal} {Phys. Rev. Lett.}\ }\textbf {\bibinfo {volume} {110}},\ \bibinfo
  {pages} {138302} (\bibinfo {year} {2013})}\BibitemShut {NoStop}%
\bibitem [{\citenamefont {Joe}\ \emph {et~al.}(2016)\citenamefont {Joe},
  \citenamefont {O'Neil}, \citenamefont {{Miaja-Avila}and J~W~Fowler},
  \citenamefont {Jimenez}, \citenamefont {Silverman}, \citenamefont {Swetz},\
  and\ \citenamefont {Ullom}}]{joe2016}%
  \BibitemOpen
  \bibfield  {author} {\bibinfo {author} {\bibfnamefont {Y.~I.}\ \bibnamefont
  {Joe}}, \bibinfo {author} {\bibfnamefont {G.~C.}\ \bibnamefont {O'Neil}},
  \bibinfo {author} {\bibfnamefont {L.}~\bibnamefont {{Miaja-Avila}and
  J~W~Fowler}}, \bibinfo {author} {\bibfnamefont {R.}~\bibnamefont {Jimenez}},
  \bibinfo {author} {\bibfnamefont {K.~L.}\ \bibnamefont {Silverman}}, \bibinfo
  {author} {\bibfnamefont {D.~S.}\ \bibnamefont {Swetz}}, \ and\ \bibinfo
  {author} {\bibfnamefont {J.~N.}\ \bibnamefont {Ullom}},\ }\href {\doibase
  10.1088/0953-4075/49/2/024003} {\bibfield  {journal} {\bibinfo  {journal}
  {Journal of Physics B: Atomic, Molecular and Optical Physics}\ }\textbf
  {\bibinfo {volume} {49}},\ \bibinfo {pages} {024003} (\bibinfo {year}
  {2016})}\BibitemShut {NoStop}%
\bibitem [{\citenamefont {Szlachetko}\ \emph {et~al.}(2012)\citenamefont
  {Szlachetko}, \citenamefont {Nachtegaal}, \citenamefont {de~Boni},
  \citenamefont {Willimann}, \citenamefont {Safonova}, \citenamefont {Sa},
  \citenamefont {Smolentsev}, \citenamefont {Szlachetko}, \citenamefont {van
  Bokhoven}, \citenamefont {Dousse}, \citenamefont {Hoszowska}, \citenamefont
  {Kayser}, \citenamefont {Jagodzinski}, \citenamefont {Bergamaschi},
  \citenamefont {Schmitt}, \citenamefont {David},\ and\ \citenamefont
  {L\"{u}cke}}]{szlachetko2012rsi}%
  \BibitemOpen
  \bibfield  {author} {\bibinfo {author} {\bibfnamefont {J.}~\bibnamefont
  {Szlachetko}}, \bibinfo {author} {\bibfnamefont {M.}~\bibnamefont
  {Nachtegaal}}, \bibinfo {author} {\bibfnamefont {E.}~\bibnamefont {de~Boni}},
  \bibinfo {author} {\bibfnamefont {M.}~\bibnamefont {Willimann}}, \bibinfo
  {author} {\bibfnamefont {O.}~\bibnamefont {Safonova}}, \bibinfo {author}
  {\bibfnamefont {J.}~\bibnamefont {Sa}}, \bibinfo {author} {\bibfnamefont
  {G.}~\bibnamefont {Smolentsev}}, \bibinfo {author} {\bibfnamefont
  {M.}~\bibnamefont {Szlachetko}}, \bibinfo {author} {\bibfnamefont {J.~A.}\
  \bibnamefont {van Bokhoven}}, \bibinfo {author} {\bibfnamefont {J.-C.}\
  \bibnamefont {Dousse}}, \bibinfo {author} {\bibfnamefont {J.}~\bibnamefont
  {Hoszowska}}, \bibinfo {author} {\bibfnamefont {Y.}~\bibnamefont {Kayser}},
  \bibinfo {author} {\bibfnamefont {P.}~\bibnamefont {Jagodzinski}}, \bibinfo
  {author} {\bibfnamefont {A.}~\bibnamefont {Bergamaschi}}, \bibinfo {author}
  {\bibfnamefont {B.}~\bibnamefont {Schmitt}}, \bibinfo {author} {\bibfnamefont
  {C.}~\bibnamefont {David}}, \ and\ \bibinfo {author} {\bibfnamefont
  {A.}~\bibnamefont {L\"{u}cke}},\ }\href {\doibase 10.1063/1.4756691}
  {\bibfield  {journal} {\bibinfo  {journal} {Review of Scientific
  Instruments}\ }\textbf {\bibinfo {volume} {83}},\ \bibinfo {eid} {103105}
  (\bibinfo {year} {2012})}\BibitemShut {NoStop}%
\bibitem [{\citenamefont {Alonso~Mori}\ \emph {et~al.}(2012)\citenamefont
  {Alonso~Mori}, \citenamefont {Kern}, \citenamefont {Sokaras}, \citenamefont
  {Weng}, \citenamefont {Nordlund}, \citenamefont {Tran}, \citenamefont
  {Montanez}, \citenamefont {Delor}, \citenamefont {Yachandra}, \citenamefont
  {Yano},\ and\ \citenamefont {Bergmann}}]{alonsomori2012}%
  \BibitemOpen
  \bibfield  {author} {\bibinfo {author} {\bibfnamefont {R.}~\bibnamefont
  {Alonso~Mori}}, \bibinfo {author} {\bibfnamefont {J.}~\bibnamefont {Kern}},
  \bibinfo {author} {\bibfnamefont {D.}~\bibnamefont {Sokaras}}, \bibinfo
  {author} {\bibfnamefont {T.-C.}\ \bibnamefont {Weng}}, \bibinfo {author}
  {\bibfnamefont {D.}~\bibnamefont {Nordlund}}, \bibinfo {author}
  {\bibfnamefont {R.}~\bibnamefont {Tran}}, \bibinfo {author} {\bibfnamefont
  {P.}~\bibnamefont {Montanez}}, \bibinfo {author} {\bibfnamefont
  {J.}~\bibnamefont {Delor}}, \bibinfo {author} {\bibfnamefont {V.~K.}\
  \bibnamefont {Yachandra}}, \bibinfo {author} {\bibfnamefont {J.}~\bibnamefont
  {Yano}}, \ and\ \bibinfo {author} {\bibfnamefont {U.}~\bibnamefont
  {Bergmann}},\ }\href {\doibase 10.1063/1.4737630} {\bibfield  {journal}
  {\bibinfo  {journal} {Review of Scientific Instruments}\ }\textbf {\bibinfo
  {volume} {83}},\ \bibinfo {eid} {073114} (\bibinfo {year}
  {2012})}\BibitemShut {NoStop}%
\bibitem [{\citenamefont {Hoszowska}\ and\ \citenamefont
  {Dousse}(2004)}]{hoszowska2004}%
  \BibitemOpen
  \bibfield  {author} {\bibinfo {author} {\bibfnamefont {J.}~\bibnamefont
  {Hoszowska}}\ and\ \bibinfo {author} {\bibfnamefont {J.-C.}\ \bibnamefont
  {Dousse}},\ }\href {\doibase 10.1016/j.elspec.2004.02.005} {\bibfield
  {journal} {\bibinfo  {journal} {Journal of Electron Spectroscopy and Related
  Phenomena}\ }\textbf {\bibinfo {volume} {137–140}},\ \bibinfo {pages} {687
  } (\bibinfo {year} {2004})}\BibitemShut {NoStop}%
\bibitem [{\citenamefont {Lecante}\ \emph {et~al.}(1994)\citenamefont
  {Lecante}, \citenamefont {Jaud}, \citenamefont {Mosset}, \citenamefont
  {Galy},\ and\ \citenamefont {Burian}}]{Lecante1994}%
  \BibitemOpen
  \bibfield  {author} {\bibinfo {author} {\bibfnamefont {P.}~\bibnamefont
  {Lecante}}, \bibinfo {author} {\bibfnamefont {J.}~\bibnamefont {Jaud}},
  \bibinfo {author} {\bibfnamefont {A.}~\bibnamefont {Mosset}}, \bibinfo
  {author} {\bibfnamefont {J.}~\bibnamefont {Galy}}, \ and\ \bibinfo {author}
  {\bibfnamefont {A.}~\bibnamefont {Burian}},\ }\href {\doibase
  http://dx.doi.org/10.1063/1.1144909} {\bibfield  {journal} {\bibinfo
  {journal} {Review of Scientific Instruments}\ }\textbf {\bibinfo {volume}
  {65}},\ \bibinfo {pages} {845} (\bibinfo {year} {1994})}\BibitemShut
  {NoStop}%
\bibitem [{\citenamefont {Inada}, \citenamefont {Funahashi},\ and\
  \citenamefont {Ohtaki}(1994)}]{Inada1994}%
  \BibitemOpen
  \bibfield  {author} {\bibinfo {author} {\bibfnamefont {Y.}~\bibnamefont
  {Inada}}, \bibinfo {author} {\bibfnamefont {S.}~\bibnamefont {Funahashi}}, \
  and\ \bibinfo {author} {\bibfnamefont {H.}~\bibnamefont {Ohtaki}},\ }\href
  {\doibase http://dx.doi.org/10.1063/1.1144775} {\bibfield  {journal}
  {\bibinfo  {journal} {Review of Scientific Instruments}\ }\textbf {\bibinfo
  {volume} {65}},\ \bibinfo {pages} {18} (\bibinfo {year} {1994})}\BibitemShut
  {NoStop}%
\bibitem [{\citenamefont {Schlesiger}\ \emph {et~al.}(2015)\citenamefont
  {Schlesiger}, \citenamefont {Anklamm}, \citenamefont {Stiel}, \citenamefont
  {Malzer},\ and\ \citenamefont {Kanngie{\ss}er}}]{Schlesiger2015}%
  \BibitemOpen
  \bibfield  {author} {\bibinfo {author} {\bibfnamefont {C.}~\bibnamefont
  {Schlesiger}}, \bibinfo {author} {\bibfnamefont {L.}~\bibnamefont {Anklamm}},
  \bibinfo {author} {\bibfnamefont {H.}~\bibnamefont {Stiel}}, \bibinfo
  {author} {\bibfnamefont {W.}~\bibnamefont {Malzer}}, \ and\ \bibinfo {author}
  {\bibfnamefont {B.}~\bibnamefont {Kanngie{\ss}er}},\ }\href {\doibase
  10.1039/C4JA00303A} {\bibfield  {journal} {\bibinfo  {journal} {J. Anal. At.
  Spectrom.}\ }\textbf {\bibinfo {volume} {30}},\ \bibinfo {pages} {1080}
  (\bibinfo {year} {2015})}\BibitemShut {NoStop}%
\bibitem [{\citenamefont {Kayser}\ \emph {et~al.}(2014)\citenamefont {Kayser},
  \citenamefont {B{\l{}}achucki}, \citenamefont {Dousse}, \citenamefont
  {Hoszowska}, \citenamefont {Neff},\ and\ \citenamefont
  {Romano}}]{Kayser2014}%
  \BibitemOpen
  \bibfield  {author} {\bibinfo {author} {\bibfnamefont {Y.}~\bibnamefont
  {Kayser}}, \bibinfo {author} {\bibfnamefont {W.}~\bibnamefont
  {B{\l{}}achucki}}, \bibinfo {author} {\bibfnamefont {J.-C.}\ \bibnamefont
  {Dousse}}, \bibinfo {author} {\bibfnamefont {J.}~\bibnamefont {Hoszowska}},
  \bibinfo {author} {\bibfnamefont {M.}~\bibnamefont {Neff}}, \ and\ \bibinfo
  {author} {\bibfnamefont {V.}~\bibnamefont {Romano}},\ }\href {\doibase
  10.1063/1.4869340} {\bibfield  {journal} {\bibinfo  {journal} {Review of
  Scientific Instruments}\ }\textbf {\bibinfo {volume} {85}},\ \bibinfo {eid}
  {043101} (\bibinfo {year} {2014})}\BibitemShut {NoStop}%
\bibitem [{\citenamefont {Vank{\'{o}}}\ \emph {et~al.}(2010)\citenamefont
  {Vank{\'{o}}}, \citenamefont {Glatzel}, \citenamefont {Pham}, \citenamefont
  {Abela}, \citenamefont {Grolimund}, \citenamefont {Borca}, \citenamefont
  {Johnson}, \citenamefont {Milne},\ and\ \citenamefont
  {Bressler}}]{vanko2010}%
  \BibitemOpen
  \bibfield  {author} {\bibinfo {author} {\bibfnamefont {G.}~\bibnamefont
  {Vank{\'{o}}}}, \bibinfo {author} {\bibfnamefont {P.}~\bibnamefont
  {Glatzel}}, \bibinfo {author} {\bibfnamefont {V.-T.}\ \bibnamefont {Pham}},
  \bibinfo {author} {\bibfnamefont {R.}~\bibnamefont {Abela}}, \bibinfo
  {author} {\bibfnamefont {D.}~\bibnamefont {Grolimund}}, \bibinfo {author}
  {\bibfnamefont {C.~N.}\ \bibnamefont {Borca}}, \bibinfo {author}
  {\bibfnamefont {S.~L.}\ \bibnamefont {Johnson}}, \bibinfo {author}
  {\bibfnamefont {C.~J.}\ \bibnamefont {Milne}}, \ and\ \bibinfo {author}
  {\bibfnamefont {C.}~\bibnamefont {Bressler}},\ }\href {\doibase
  10.1002/anie.201000844} {\bibfield  {journal} {\bibinfo  {journal}
  {Angewandte Chemie International Edition}\ }\textbf {\bibinfo {volume}
  {49}},\ \bibinfo {pages} {5910} (\bibinfo {year} {2010})}\BibitemShut
  {NoStop}%
\bibitem [{\citenamefont {March}\ \emph {et~al.}(2015)\citenamefont {March},
  \citenamefont {Assefa}, \citenamefont {Bressler}, \citenamefont {Doumy},
  \citenamefont {Galler}, \citenamefont {Gawelda}, \citenamefont {Kanter},
  \citenamefont {NĂ©meth}, \citenamefont {PĂˇpai}, \citenamefont
  {Southworth}, \citenamefont {Young},\ and\ \citenamefont
  {VankĂł}}]{March2015}%
  \BibitemOpen
  \bibfield  {author} {\bibinfo {author} {\bibfnamefont {A.~M.}\ \bibnamefont
  {March}}, \bibinfo {author} {\bibfnamefont {T.~A.}\ \bibnamefont {Assefa}},
  \bibinfo {author} {\bibfnamefont {C.}~\bibnamefont {Bressler}}, \bibinfo
  {author} {\bibfnamefont {G.}~\bibnamefont {Doumy}}, \bibinfo {author}
  {\bibfnamefont {A.}~\bibnamefont {Galler}}, \bibinfo {author} {\bibfnamefont
  {W.}~\bibnamefont {Gawelda}}, \bibinfo {author} {\bibfnamefont {E.~P.}\
  \bibnamefont {Kanter}}, \bibinfo {author} {\bibfnamefont {Z.}~\bibnamefont
  {NĂ©meth}}, \bibinfo {author} {\bibfnamefont {M.}~\bibnamefont {PĂˇpai}},
  \bibinfo {author} {\bibfnamefont {S.~H.}\ \bibnamefont {Southworth}},
  \bibinfo {author} {\bibfnamefont {L.}~\bibnamefont {Young}}, \ and\ \bibinfo
  {author} {\bibfnamefont {G.}~\bibnamefont {VankĂł}},\ }\href {\doibase
  10.1021/jp511838q} {\bibfield  {journal} {\bibinfo  {journal} {The Journal of
  Physical Chemistry C}\ }\textbf {\bibinfo {volume} {119}},\ \bibinfo {pages}
  {14571} (\bibinfo {year} {2015})},\ \Eprint
  {http://arxiv.org/abs/http://dx.doi.org/10.1021/jp511838q}
  {http://dx.doi.org/10.1021/jp511838q} \BibitemShut {NoStop}%
\bibitem [{\citenamefont {Alain}\ \emph {et~al.}(2009)\citenamefont {Alain},
  \citenamefont {Jacques}, \citenamefont {Diane},\ and\ \citenamefont
  {Karine}}]{Alain2009}%
  \BibitemOpen
  \bibfield  {author} {\bibinfo {author} {\bibfnamefont {M.}~\bibnamefont
  {Alain}}, \bibinfo {author} {\bibfnamefont {M.}~\bibnamefont {Jacques}},
  \bibinfo {author} {\bibfnamefont {M.-B.}\ \bibnamefont {Diane}}, \ and\
  \bibinfo {author} {\bibfnamefont {P.}~\bibnamefont {Karine}},\ }\href
  {http://stacks.iop.org/1742-6596/190/i=1/a=012034} {\bibfield  {journal}
  {\bibinfo  {journal} {Journal of Physics: Conference Series}\ }\textbf
  {\bibinfo {volume} {190}},\ \bibinfo {pages} {012034} (\bibinfo {year}
  {2009})}\BibitemShut {NoStop}%
\bibitem [{\citenamefont {Pan}\ \emph {et~al.}(2012)\citenamefont {Pan},
  \citenamefont {Jian}, \citenamefont {Ablat}, \citenamefont {Li},
  \citenamefont {Sun},\ and\ \citenamefont {Wu}}]{Pan2012}%
  \BibitemOpen
  \bibfield  {author} {\bibinfo {author} {\bibfnamefont {D.}~\bibnamefont
  {Pan}}, \bibinfo {author} {\bibfnamefont {J.~K.}\ \bibnamefont {Jian}},
  \bibinfo {author} {\bibfnamefont {A.}~\bibnamefont {Ablat}}, \bibinfo
  {author} {\bibfnamefont {J.}~\bibnamefont {Li}}, \bibinfo {author}
  {\bibfnamefont {Y.~F.}\ \bibnamefont {Sun}}, \ and\ \bibinfo {author}
  {\bibfnamefont {R.}~\bibnamefont {Wu}},\ }\href {\doibase
  http://dx.doi.org/10.1063/1.4749408} {\bibfield  {journal} {\bibinfo
  {journal} {Journal of Applied Physics}\ }\textbf {\bibinfo {volume} {112}},\
  \bibinfo {eid} {053911} (\bibinfo {year} {2012}),\
  http://dx.doi.org/10.1063/1.4749408}\BibitemShut {NoStop}%
\bibitem [{\citenamefont {Glatzel}, \citenamefont {Smolentsev},\ and\
  \citenamefont {Bunker}(2009)}]{glatzel2009xafs14}%
  \BibitemOpen
  \bibfield  {author} {\bibinfo {author} {\bibfnamefont {P.}~\bibnamefont
  {Glatzel}}, \bibinfo {author} {\bibfnamefont {G.}~\bibnamefont {Smolentsev}},
  \ and\ \bibinfo {author} {\bibfnamefont {G.}~\bibnamefont {Bunker}},\ }\href
  {\doibase 10.1088/1742-6596/190/1/012046} {\bibfield  {journal} {\bibinfo
  {journal} {Journal of Physics: Conference Series}\ }\textbf {\bibinfo
  {volume} {190}},\ \bibinfo {pages} {012046} (\bibinfo {year}
  {2009})}\BibitemShut {NoStop}%
\bibitem [{\citenamefont {Mori}, \citenamefont {Taga},\ and\ \citenamefont
  {Yamashita}(2015)}]{Mori2015}%
  \BibitemOpen
  \bibfield  {author} {\bibinfo {author} {\bibfnamefont {K.}~\bibnamefont
  {Mori}}, \bibinfo {author} {\bibfnamefont {T.}~\bibnamefont {Taga}}, \ and\
  \bibinfo {author} {\bibfnamefont {H.}~\bibnamefont {Yamashita}},\ }\href
  {\doibase 10.1002/cctc.201500101} {\bibfield  {journal} {\bibinfo  {journal}
  {ChemCatChem}\ }\textbf {\bibinfo {volume} {7}},\ \bibinfo {pages} {1285}
  (\bibinfo {year} {2015})}\BibitemShut {NoStop}%
\bibitem [{\citenamefont {Tirez}\ \emph {et~al.}(2011)\citenamefont {Tirez},
  \citenamefont {Silversmit}, \citenamefont {Vincze}, \citenamefont {Servaes},
  \citenamefont {Vanhoof}, \citenamefont {Mertens}, \citenamefont {Bleux},\
  and\ \citenamefont {Berghmans}}]{Tirez2011}%
  \BibitemOpen
  \bibfield  {author} {\bibinfo {author} {\bibfnamefont {K.}~\bibnamefont
  {Tirez}}, \bibinfo {author} {\bibfnamefont {G.}~\bibnamefont {Silversmit}},
  \bibinfo {author} {\bibfnamefont {L.}~\bibnamefont {Vincze}}, \bibinfo
  {author} {\bibfnamefont {K.}~\bibnamefont {Servaes}}, \bibinfo {author}
  {\bibfnamefont {C.}~\bibnamefont {Vanhoof}}, \bibinfo {author} {\bibfnamefont
  {M.}~\bibnamefont {Mertens}}, \bibinfo {author} {\bibfnamefont
  {N.}~\bibnamefont {Bleux}}, \ and\ \bibinfo {author} {\bibfnamefont
  {P.}~\bibnamefont {Berghmans}},\ }\href {\doibase 10.1039/C0JA00049C}
  {\bibfield  {journal} {\bibinfo  {journal} {J. Anal. At. Spectrom.}\ }\textbf
  {\bibinfo {volume} {26}},\ \bibinfo {pages} {517} (\bibinfo {year}
  {2011})}\BibitemShut {NoStop}%
\bibitem [{\citenamefont {Medarde}\ \emph {et~al.}(2009)\citenamefont
  {Medarde}, \citenamefont {Dallera}, \citenamefont {Grioni}, \citenamefont
  {Delley}, \citenamefont {Vernay}, \citenamefont {Mesot}, \citenamefont
  {Sikora}, \citenamefont {Alonso},\ and\ \citenamefont
  {Mart\'{\i}nez-Lope}}]{Medarde2009}%
  \BibitemOpen
  \bibfield  {author} {\bibinfo {author} {\bibfnamefont {M.}~\bibnamefont
  {Medarde}}, \bibinfo {author} {\bibfnamefont {C.}~\bibnamefont {Dallera}},
  \bibinfo {author} {\bibfnamefont {M.}~\bibnamefont {Grioni}}, \bibinfo
  {author} {\bibfnamefont {B.}~\bibnamefont {Delley}}, \bibinfo {author}
  {\bibfnamefont {F.}~\bibnamefont {Vernay}}, \bibinfo {author} {\bibfnamefont
  {J.}~\bibnamefont {Mesot}}, \bibinfo {author} {\bibfnamefont
  {M.}~\bibnamefont {Sikora}}, \bibinfo {author} {\bibfnamefont {J.~A.}\
  \bibnamefont {Alonso}}, \ and\ \bibinfo {author} {\bibfnamefont {M.~J.}\
  \bibnamefont {Mart\'{\i}nez-Lope}},\ }\href {\doibase
  10.1103/PhysRevB.80.245105} {\bibfield  {journal} {\bibinfo  {journal} {Phys.
  Rev. B}\ }\textbf {\bibinfo {volume} {80}},\ \bibinfo {pages} {245105}
  (\bibinfo {year} {2009})}\BibitemShut {NoStop}%
\bibitem [{\citenamefont {Zhang}\ \emph {et~al.}(2015)\citenamefont {Zhang},
  \citenamefont {Brugger}, \citenamefont {Etschmann}, \citenamefont {Ngothai},\
  and\ \citenamefont {Zeng}}]{Zhang2015}%
  \BibitemOpen
  \bibfield  {author} {\bibinfo {author} {\bibfnamefont {N.}~\bibnamefont
  {Zhang}}, \bibinfo {author} {\bibfnamefont {J.}~\bibnamefont {Brugger}},
  \bibinfo {author} {\bibfnamefont {B.}~\bibnamefont {Etschmann}}, \bibinfo
  {author} {\bibfnamefont {Y.}~\bibnamefont {Ngothai}}, \ and\ \bibinfo
  {author} {\bibfnamefont {D.}~\bibnamefont {Zeng}},\ }\href {\doibase
  10.1371/journal.pone.0119805} {\bibfield  {journal} {\bibinfo  {journal}
  {PLoS ONE}\ }\textbf {\bibinfo {volume} {10}},\ \bibinfo {pages} {1}
  (\bibinfo {year} {2015})}\BibitemShut {NoStop}%
\bibitem [{\citenamefont {Woolley}\ \emph {et~al.}(2011)\citenamefont
  {Woolley}, \citenamefont {Illy}, \citenamefont {Ryan},\ and\ \citenamefont
  {Skinner}}]{woolley2011}%
  \BibitemOpen
  \bibfield  {author} {\bibinfo {author} {\bibfnamefont {R.~J.}\ \bibnamefont
  {Woolley}}, \bibinfo {author} {\bibfnamefont {B.~N.}\ \bibnamefont {Illy}},
  \bibinfo {author} {\bibfnamefont {M.~P.}\ \bibnamefont {Ryan}}, \ and\
  \bibinfo {author} {\bibfnamefont {S.~J.}\ \bibnamefont {Skinner}},\ }\href
  {\doibase 10.1039/C1JM14320D} {\bibfield  {journal} {\bibinfo  {journal} {J.
  Mater. Chem.}\ }\textbf {\bibinfo {volume} {21}},\ \bibinfo {pages} {18592}
  (\bibinfo {year} {2011})}\BibitemShut {NoStop}%
\bibitem [{\citenamefont {O'Grady}\ \emph {et~al.}(1996)\citenamefont
  {O'Grady}, \citenamefont {Pandya}, \citenamefont {Swider},\ and\
  \citenamefont {Corrigan}}]{ogrady1996}%
  \BibitemOpen
  \bibfield  {author} {\bibinfo {author} {\bibfnamefont {W.~E.}\ \bibnamefont
  {O'Grady}}, \bibinfo {author} {\bibfnamefont {K.~I.}\ \bibnamefont {Pandya}},
  \bibinfo {author} {\bibfnamefont {K.~E.}\ \bibnamefont {Swider}}, \ and\
  \bibinfo {author} {\bibfnamefont {D.~A.}\ \bibnamefont {Corrigan}},\ }\href
  {\doibase 10.1149/1.1836687} {\bibfield  {journal} {\bibinfo  {journal}
  {Journal of The Electrochemical Society}\ }\textbf {\bibinfo {volume}
  {143}},\ \bibinfo {pages} {1613} (\bibinfo {year} {1996})}\BibitemShut
  {NoStop}%
\bibitem [{\citenamefont {Vank{\'o}}\ \emph {et~al.}(2006)\citenamefont
  {Vank{\'o}}, \citenamefont {Neisius}, \citenamefont {Moln{\'a}r},
  \citenamefont {Renz}, \citenamefont {K{\'a}rp{\'a}ti}, \citenamefont
  {Shukla},\ and\ \citenamefont {de~Groot}}]{vanko2006jpcb}%
  \BibitemOpen
  \bibfield  {author} {\bibinfo {author} {\bibfnamefont {G.}~\bibnamefont
  {Vank{\'o}}}, \bibinfo {author} {\bibfnamefont {T.}~\bibnamefont {Neisius}},
  \bibinfo {author} {\bibfnamefont {G.}~\bibnamefont {Moln{\'a}r}}, \bibinfo
  {author} {\bibfnamefont {F.}~\bibnamefont {Renz}}, \bibinfo {author}
  {\bibfnamefont {S.}~\bibnamefont {K{\'a}rp{\'a}ti}}, \bibinfo {author}
  {\bibfnamefont {A.}~\bibnamefont {Shukla}}, \ and\ \bibinfo {author}
  {\bibfnamefont {F.~M.~F.}\ \bibnamefont {de~Groot}},\ }\href {\doibase
  10.1021/jp0615961} {\bibfield  {journal} {\bibinfo  {journal} {J. Phys. Chem.
  B}\ }\textbf {\bibinfo {volume} {110}},\ \bibinfo {pages} {11647} (\bibinfo
  {year} {2006})}\BibitemShut {NoStop}%
\end{thebibliography}%

\end{document}